\documentclass[%
 reprint,twocolumn,aps,pre,superscriptaddress,
 amsmath,amssymb,aps,longbibliography,
floatfix,
]{revtex4-2}
\usepackage{color} 
\usepackage[usenames,dvipsnames]{xcolor}
\definecolor{darkblue}{rgb}{0,0,0.6}
\usepackage{graphicx}
\usepackage{dcolumn}
\usepackage{bm}
\usepackage{physics, bm,bbm, dsfont}
\usepackage{mathrsfs}  
\usepackage{amssymb}
\usepackage{amsmath, amssymb, amscd, amsthm, amsfonts}
\usepackage{xparse}
\usepackage[colorlinks,linkcolor=darkblue,citecolor=darkblue,urlcolor=darkblue]{hyperref}

\newcommand{\bxi}{\boldsymbol{\xi}}

\newcommand{\bz}{\text{\bf z}}

\newcommand{\ee}{\text{e}}

\newcommand{\p}{\partial}
\newcommand{\bx}{\text{\bf x}}
\newcommand{\bff}{\text{\bf f}}
\newcommand{\br}{\text{\bf r}}
\newcommand{\be}{\text{\bf e}}
\newcommand{\eps}{\varepsilon}
\newcommand{\bk}{\text{\bf k}}

\newcommand{\bq}{\text{\bf q}}
\newcommand{\bp}{\text{\bf p}}

\newcommand{\bu}{\text{\bf u}}

\newcommand{\bj}{{\bf j}}

\newcommand{\bv}{\text{\bf v}}

\newcommand{\bchi}{\boldsymbol{\chi}}

\newcommand{\bXi}{\boldsymbol{\Xi}}
\newcommand{\bnabla}{\boldsymbol{\nabla}}
\newcommand{\bmu}{\boldsymbol{\mu}}

\newcommand{\bn}{\text{\bf n}}

\newcommand{\bF}{\text{\bf F}}
\newcommand{\bG}{\text{\bf G}}

\newcommand{\bA}{\text{\bf A}}
\newcommand{\bD}{\text{\bf D}}
\newcommand{\bO}{\text{\bf O}}
\newcommand{\bK}{\text{\bf K}}
\newcommand{\bV}{\text{\bf V}}
\newcommand{\bU}{\text{\bf U}}
\newcommand{\bR}{\text{\bf R}}
\newcommand{\bJ}{\text{\bf J}}

\newcommand{\bqhat}{\boldsymbol{\widehat{\textbf{q}}}}
\newcommand{\buhat}{\boldsymbol{\widehat{\textbf{u}}}}
\newcommand{\bkhat}{\boldsymbol{\hat{\textbf{k}}}}
\newcommand{\bFext}{\bF_\text{ext}}
\newcommand{\khat}{\hat{k}}
\newcommand{\mQ}{\mathcal{Q}}
\newcommand{\mP}{\mathcal{P}}
\newcommand{\mId}{\mathcal{I}}
\newcommand{\bMtilde}{\mathbf{\widetilde{\text{ \bf M}}}}
\newcommand{\bKtilde}{\mathbf{\widetilde{\text{ \bf K}}}}
\newcommand{\mR}{\mathcal{R}}
\newcommand{\Rotpihalf}{\mathcal{R}_{\frac{\pi}{2}}}

\newcommand{\nt}{n_0}
\newcommand{\Omegat}{\Omega_{\gamma,\text{ext}}} 
\newcommand{\Omegairr}{\Omega_{{-\gamma}}^{\text{irrs}}}

\newcommand{\phif}{\phi_{\text{\tiny{f}}}}

\ExplSyntaxOn
\NewDocumentCommand{\eqrefs}{m}
 {
  \joansola_eqrefs:n { #1 }
 }

\seq_new:N \l__joansola_eqrefs_in_seq
\seq_new:N \l__joansola_eqrefs_out_seq

\cs_new_protected:Nn \joansola_eqrefs:n
 {
  \seq_set_split:Nnn \l__joansola_eqrefs_in_seq { , } { #1 }
  \seq_set_map:NNn \l__joansola_eqrefs_out_seq \l__joansola_eqrefs_in_seq
   {
    \exp_not:n { \ref{##1} }
   }
  \int_compare:nTF { \seq_count:N \l__joansola_eqrefs_out_seq < 2 }
   { Eq.\nobreakspace }
   { Eqs.\nobreakspace }
  \textup{(\seq_use:Nn \l__joansola_eqrefs_out_seq {,~})}
 }
\ExplSyntaxOff

\begin{document}

\title{Irreversible Boltzmann samplers in dense liquids: weak-coupling approximation and mode-coupling theory}

\author{Federico Ghimenti}

\affiliation{Laboratoire Mati\`ere et Syst\`emes Complexes (MSC), Université Paris Cité  \& CNRS (UMR 7057), 75013 Paris, France}

\author{Ludovic Berthier}

\affiliation{Laboratoire Charles Coulomb (L2C), Université de Montpellier \& CNRS (UMR 5221), 34095 Montpellier, France}

\affiliation{Gulliver, UMR CNRS 7083, ESPCI Paris, PSL Research University, 75005 Paris, France}

\author{Grzegorz Szamel}

\affiliation{Department of Chemistry, Colorado State University, Fort Collins, Colorado 80523, United States of America}

\author{Fr\'ed\'eric van Wijland}

\affiliation{Laboratoire Mati\`ere et Syst\`emes Complexes (MSC), Université Paris Cité  \& CNRS (UMR 7057), 75013 Paris, France}

\date{\today}

\begin{abstract}
Exerting a nonequilibrium drive on an otherwise equilibrium Langevin process brings the dynamics out of equilibrium but can also speedup the approach to the Boltzmann steady-state. Transverse forces are a minimal framework to achieve dynamical acceleration of the Boltzmann sampling. We consider a simple liquid in three space dimensions subjected to additional transverse pairwise forces, and quantify the extent to which transverse forces accelerate the dynamics. We first explore the dynamics of a tracer in a weak coupling regime describing high temperatures. The resulting acceleration is correlated with a monotonous increase of the magnitude of odd transport coefficients (mobility and diffusivity) with the amplitude of the transverse drive. We then develop a nonequilibrium version of the mode-coupling theory able to capture the effect of transverse forces, and more generally of forces created by additional degrees of freedom. Based on an analysis of transport coefficients, both odd and longitudinal, both for the collective modes and for a tracer particle, we find a systematic acceleration of the dynamics. Quantitatively, the gain, which is guaranteed throughout the ergodic phase, turns out to be a decreasing function of temperature beyond a temperature crossover, in particular as the glass transition is approached. Our theoretical results are in good agreement with available numerical results.
\end{abstract}

\maketitle

\section{Introduction}

When interested in the dynamics of dense liquids~\cite{berthier2011theoretical}, the mode-coupling approximation~\cite{gotze1999recent,gotze2009complex} is a method of choice. It is a versatile tool designed to analyze the dynamics of simple liquids and of other soft-matter systems in realistic space dimensions ($d=2 $ or $3$). It has famously been applied to critical dynamics~\cite{kawasaki1976renormalization}, to glass-formers~\cite{gotze2009complex}, to polymers~\cite{schweizer1989microscopic} and colloidal assemblies~\cite{Klein2002,pham2002multiple} up to hard-condensed matter systems~\cite{shah2012hot}. These systems share the common trait that they are in thermal equilibrium. In recent years, the mode-coupling approximation has been extended to some nonequilibrium settings, such as sheared liquids~\cite{fuchs2009mode}, granular fluids~\cite{hayakawa2008mode, kranz2010glass}, or, even more recently, to systems of self-propelled particles~\cite{szamel2016theory,liluashvili2017mode,szamel2019mode,debets2023mode}. Here, our goal is to use these works as an inspiration to develop a mode-coupling approximation for systems of particles driven by nonequilibrium forces that nevertheless sample the Boltzmann distribution in their steady-state. Such systems are of special interest because, for specifically designed nonequilibrium forces, they can sample the Boltzmann distribution faster than their reversible equilibrium counterparts. This is of course an interesting property when dynamical evolution is intrinsically sluggish, as in dense and cold liquids~\cite{berthier2023modern}.

When confronted with the problem of sampling the Boltzmann distribution for a system of interacting particles, a practitioner of simulations can resort to several standard techniques~\cite{frenkel2001understanding,allen2017computer} such as Metropolis-Hastings Monte Carlo, Brownian and Newtonian Dynamics. These techniques endow the system with reversible dynamics, ensuring that at long times the equilibrium Boltzmann distribution is sampled. It is a practical challenge to reduce as much as possible the time required to achieve convergence to the equilibrium distribution. This convergence time proves overwhelmingly long for many systems of interest, from proteins to supercooled liquids and neural networks. Recently, several techniques were proposed to accelerate the dynamics~\cite{berthier2023modern}, including the very efficient swap Monte Carlo algorithm that achieves dramatic speedup for size polydisperse liquids~\cite{berthier2016equilibrium,ninarello2017models}.

An alternative approach, which has recently garnered attention, is to exploit an irreversible dynamics, namely one that does not respect detailed balance, but that at the same time ensures that the Boltzmann distribution is properly sampled. The idea is that by going out-of-equilibrium a faster exploration of the phase space is sometimes possible. To achieve this goal one could add, for example, a drift perpendicular to the gradient flow of the system, allowing for an out of equilibrium dynamics with the aforementioned property. We will refer to such an orthogonal drift as transverse forces. The possibility of using transverse forces was first envisaged in \cite{hwang1993accelerating}, where it was also shown that the presence of a nonequilibrium drive increases the spectral gap of the associated dynamical evolution operator, allowing for shorter convergence time. A similar proof was later discussed in \cite{ichiki2013violation} in the context of Markov Chains. More detailed analysis of the exact amount of the acceleration were carried out in the case of a particle in a harmonic well~\cite{hwang2005accelerating, lelievre2013optimal}. Generic inequalities concerning asymptotic variance (a projection of the convergence time)~\cite{hwang2015variance,duncan2016variance} and large deviation functions~\cite{rey2015irreversible, kaiser2017acceleration, coghi2021role} were also obtained, all demonstrating a better performance of the transverse force dynamics with respect to its equilibrium counterpart, in agreement with existing mathematical theorems.

The acceleration cannot however increase indefinitely. In \cite{franke2010behavior} it was shown that the spectral gap saturates in the limit of a transverse force with infinite amplitude, although little is known about this actual limiting value. As a matter of fact, despite the abundance of mathematical inequalities and rigorous bounds, there exist very few quantitative results regarding the speedup actually obtained in systems of theoretical and practical interest. Efforts in this direction consist in some studies on benchmark neural network problems~\cite{Ohzeki_2016, futami2020accelerating, gao2020breaking, futami2021accelerated}, in numerical results on double well and $XY$ models~\cite{ohzeki2015langevin}, and the zero-range process~\cite{kaiser2017acceleration}. Results on the barrier crossing problem in the presence of transverse forces were also obtained~\cite{bouchet2016generalisation}. Previously, some of us studied the effect of transverse forces on the hard sampling problem of a mean-field spin glass belonging to the RFOT universality class~\cite{ghimenti2022accelerating}. In a recent Letter~\cite{ghimenti2023sampling}, we addressed the dynamics of dense liquids by a combination of numerical methods and of analytical approaches, based on the dynamical-mean-field theory and on the mode-coupling approximation. The former infinite-dimensional approach has been further detailed in \cite{ghimenti2024transverse}. The present work addresses the acceleration witnessed in finite-dimensional simulations~\cite{ghimenti2023sampling} by exploring the corresponding phenomenology through the lens of the mode-coupling approximation. We wish to quantify theoretically the acceleration provided by irreversible dynamics in a generic system of interacting particles in finite dimensions. 

After a presentation of our model in Sec.~\ref{sec:TF}, we devote Sec.~\ref{sec:RPA} to a study of the dynamics in the simpler weak coupling approximation, achieved for instance in a high-temperature regime. We use this approach to qualitatively explain the acceleration that transverse forces exert and the emergence of odd transport coefficients that help characterize these driven dynamics. In Sec.~\ref{sec:MCThardcore} we present the core of our mode-coupling approximation for transverse forces. The standard schematic approximation is shown to be missing the gist of transverse forces which leads us to present an improved version of the schematic approach. The emergence of odd transport is also discussed. In Sec.~\ref{sec:MCTtracer}, we complement the analysis of collective modes by an in-depth description of the dynamics of a tracer particle. In particular, the mismatch between the diffusivity tensor and the mobility tensor helps us shape a more intuitive picture of how acceleration operates. In \cite{ghimenti2023sampling} it was argued that the basic physics at work in the acceleration provided by transverse forces was of a similar nature as that induced by what is known in the mathematical community as lifting~\cite{diaconis2000analysis}. These two families of nonequilibrium drives are not usually treated on an equal footing. By constructing a mode-coupling approach for a lifted process, and by comparing our results with those of Sec.~\ref{sec:MCThardcore}, we are able to better substantiate our claim. Conclusions and outlook are finally gathered in Sec.~\ref{sec:outlook}, with pointers in the direction of applications to chiral active matter.

\section{Dynamics with transverse forces}

\label{sec:TF}

Our starting point is the following dynamics for a fluid of interacting particles with positions $\br_i$ in three dimensions~\cite{ghimenti2023sampling}
\begin{equation}\label{eq:DynGamma}
  \mathbf{\dot r}_i = \mu_0 \left(\mathbf{1} + \gamma \bA \right) \mathbf{F}_i + \sqrt{2\mu_0T}\bm\xi_i, 
\end{equation}
with $\mathbf{F}_i \equiv -\sum_{i} \bm\nabla_i V(\lvert\mathbf{r}_i - \mathbf{r}_j\rvert)$ a conservative force arising from a pairwise, isotropic interaction potential, and the $\bm\xi_i$'s are independent Gaussian white noises with independent components. The bare mobility $\mu_0$ and the temperature $T$ are related to the diffusion constant of a free particle $D_0$ through Einstein's relation $D_0=\mu_0 T$, setting the Boltzmann constant to unity. This dynamics differs from its overdamped equilibrium counterpart due to the presence of an antisymmetric matrix $\bA = -\bA^T$. The resulting additional transverse force $\bA \bF_i= A_{\alpha\beta}F_{i,\beta}$ injects into the system a nonequilibrium current that nevertheless preserves the Boltzmann distribution in the nonequilibrium steady-state. This means in particular that in the presence of transverse forces, the dressed mobility and diffusivity will pick up an odd component (as most other transport coefficients, such as viscosity), and we do not expect the Einstein relation to extend to the full dressed diffusivity and mobility tensors.

We would now like to take a few lines to discuss how we choose the matrix $\bA$. For concreteness, we control the strength of the nonequilibrium drive through the parameter $\gamma$ while imposing the condition that the Frobenius norm of $\bA$ is unity, namely
\begin{equation}\label{eq:frobenius}
  \sqrt{\sum_{i,j}A_{ij}^2}=1.
\end{equation}
Under this condition, the matrix $\bA$ reads
\begin{equation}\label{eq:Ageneral}
  \bA = \begin{bmatrix} 0 & A_{12} & A_{13} \\ -A_{12} & 0 & A_{23} \\ -A_{13} & -A_{23} & 0 \end{bmatrix} ,
\end{equation}
with $\sqrt{A_{12}^2 + A_{13}^2 + A_{23}^2}=1$. We now show that upon a suitable set of rotations of coordinates the matrix $\bA$ can be recast in a simpler form. Indeed, if $\bR$ is an orthogonal matrix representing a rotation in  three dimensions, i.e. $\det(\bR)=1$ and $\bR^T\bR=\mathbf{1}$, then applying $\bR$ to both sides of Eq.~\eqref{eq:DynGamma} yields
\begin{equation}\label{eq:DynGammaRot}
    \mathbf{\dot r'}_i = \mu_0 \left(\mathbf{1} + \gamma \bA' \right) \mathbf{F}'_i + \sqrt{2\mu_0T}\bm\xi'_i ,
\end{equation}
where we have defined $\br_i'\equiv \bR\br_i$, $\bF_i'\equiv \bR\bF_i$, $\bxi_i'\equiv \bR\bxi_i$ and $\bA'\equiv \bR\bA\bR^T$. The dynamics in Eq.~\eqref{eq:DynGammaRot} is the same as in Eq.~\eqref{eq:DynGamma}, in a rotated reference frame. Of course the force $\bF'_i$ is still a  central force in the rotated reference frame,
\begin{equation}
  \begin{aligned}
    F'_{i,\alpha} &= \sum_j R_{\alpha\beta}\partial_{r_{i,\beta}}V(\lvert \br_i - \br_j\rvert) \\
    &= \sum_j R_{\alpha\beta}\frac{\partial r_{i,\beta}}{\partial r'_{i,\gamma}}\partial_{r'_{i,\gamma}}V(\lvert \br_i - \br_j\rvert) \\
    &= \sum_j R_{\alpha\beta}R^T_{\beta\gamma}\partial_{r'_{i,\gamma}}V(\lvert \br'_i - \br'_j\rvert) \\
    &= \sum_j \partial_{r'_{i,\alpha}}V(\lvert \br'_i - \br'_j\rvert),
  \end{aligned}
\end{equation}
and the noises $\bxi_i'$ have the same statistics as the noises $\bxi_i$:
\begin{equation}
  \begin{aligned}
    \left\langle \xi'_{i,\alpha}(t)\xi'_{j,\beta}(t') \right\rangle &= R_{\alpha\alpha'}R_{\beta\beta'}\left\langle\xi_{i,\alpha'}(t)\xi_{j,\beta'}(t')\right\rangle \\
    &= R_{\alpha\alpha'}R_{\beta\beta'}\delta_{ij}\delta_{\alpha'\beta'}\delta(t-t') \\
    &= \delta_{ij}\delta_{\alpha\beta}\delta(t-t').
  \end{aligned}
\end{equation}
The matrix $\bA'$ is still an antisymmetric matrix satisfying Eq.~\eqref{eq:frobenius}. Without loss of generality, we can take for $\bA$ the following explicit form:
\begin{equation}\label{eq:defA}
  \bA \equiv \begin{bmatrix} 0 & -1 & 0 \\ 1 & 0 & 0 \\ 0 & 0 & 0 \end{bmatrix},
\end{equation}
which is mathematically similar to $\bA$ in Eq.~\eqref{eq:Ageneral} up to a rotation $\bR=\bR_\bx(\theta)\bR_\bz(\phi)$, with $\bR_\bx(\theta)$, $\bR_\bz(\phi)$ rotation around the $x$ and $z$ axis respectively, satisfying $\tan(\phi)=-\frac{A_{23}}{A_{13}}$, $\tan(\theta)=-\frac{1}{A_{12}}\sqrt{A^2_{13}+A_{23}^2}$. Note that for a nonzero matrix $\bA$ the dynamics is no longer isotropic. It is invariant only by rotations generated by the matrix $\bA$ in Eq.~\eqref{eq:defA}. We identify the plane left invariant by this rotations as the $(xy)$ plane.

We have now set the stage for our analysis of transverse forces. We begin with a weak-coupling approximation, which follows an approach often termed the random-phase approximation~\cite{hansen2013theory}.

\section{Weak coupling approximation}

\label{sec:RPA}

Before entering the details of the mode-coupling theory derivation, it is useful to consider the problem of a tracer diffusing in a system interacting with transverse forces in the special case in which fluctuations in the surrounding fluid can be considered as weak. Following \cite{demery2014generalized}, we  assume that the tracer produces a small perturbation in the density field of its environment, a procedure known as the random phase approximation, and derive an expression for the transport coefficients of the tracer, namely the longitudinal and the odd diffusion constants. This gives a first grasp on the speedup generated by transverse forces. In spite of throwing an interesting physical light, this approach does not easily extend to low temperatures where the weak coupling hypothesis is not fulfilled. 

\subsection{Equation of motion for a tracer}

We consider a tagged particle with position $\br_0$ diffusing in a fluid under the action of transverse forces. We include the possibility for the tagged particle to be subjected to a constant external force $\bFext$, which will be used to probe the mobility of the tagged particle. The equation of motion for the tracer is
 \begin{equation}\label{eq:eom_tracer}
    \begin{split}
        \dot\br_0 &= \mu_0\bFext - \mu_0(\mathbf{1} + \gamma\bA)\sum_{i>0} \bnabla_0 V(\br_0(t) - \br_i(t) ) \\
        &+ \sqrt{2\mu_0 T} \bxi_0 .
    \end{split}
\end{equation}
This is Eq.~\eqref{eq:DynGamma}, with the role of particle $0$ being singled out. All other particles in the fluid also interact according to Eq.~\eqref{eq:DynGamma}. We introduce the density field of the fluid without the tracer, $n(\br,t) \equiv \sum_{i>0} \delta(\br - \br_i(t))$ and we can then rewrite Eq.~\eqref{eq:eom_tracer} as
\begin{equation}\label{eq:eom_tracer_wrho}
    \begin{split}    
        \dot\br_0 &= \mu_0\bFext -(\mathbf{1} + \gamma\bA) (\bnabla V * n)(\br_0(t),t)   \\
        &+ \sqrt{2\mu_0T}\bxi_0(t).
    \end{split}
\end{equation}
with the convolution between two functions denoted by $*$: $(f*g)(\br) \equiv \int d\br f(\br-\br')g(\br')$. Equation~\eqref{eq:eom_tracer_wrho} needs to be complemented with the Dean-Kawasaki~\cite{dean1996langevin} equation for $n(\br,t)$:
\begin{equation}\label{eq:eom_rho}
    \begin{split}
        \partial_t n(\br,t) &= \mu_0 T \bnabla^2 n(\br,t) \\
        &+ \mu_0 \bnabla \cdot \left[\left(\mathbf{1} + \gamma\bA\right) n(\br,t)(\bnabla V * n)(\br,t)\right] \\
        &+ \mu_0 \bnabla \cdot \left[(\mathbf{1} + \gamma\bA)\cdot n(\br,t) \bnabla V(\br - \br_0(t)) \right] \\
        &+ \bnabla \cdot \sqrt{2T\mu_0 n(\br,t)} \bchi(\br,t) ,
    \end{split}
\end{equation}
where the Gaussian noise $\bxi(\br,t)$ has correlations $\left\langle \bxi(\br,t) \otimes \bxi(\br',t') \right\rangle = \mathbf{1}\delta(\br - \br')\delta(t - t')$. In order to solve~\eqrefs{eq:eom_tracer_wrho, eq:eom_rho} we resort to the random phase approximation.

\subsection{Random phase approximation}

In the language of the collective coordinate $n$, the random phase approximation is a simple linearization of the Dean-Kawasaki equation \eqref{eq:eom_rho} for the local and fluctuating density field $n(\br,t)=\sum_i\delta(\br-\br_i(t))$~\cite{demery2014generalized,demery2011perturbative,jardat2022diffusion,benois2023enhanced}. Assuming that the homogeneous density of the system  $\rho_0 \equiv \frac{N}{V}$ is large enough so that density fluctuations remain small, we split $n(\br,t)$ into
\begin{equation}
    n(\br,t) = \rho_0\left(1 + \frac{1}{\sqrt{\rho_0}}\phi(\br,t)\right)
\end{equation}
with $\phi(\br,t)$ a fluctuating field. To linear order in $\phi/\sqrt{\rho_0}$, which we assume to be small, Eq.~\eqref{eq:eom_rho} becomes
\begin{equation}\label{eq:phi_rpa}
    \begin{split}
         \partial_t \phi =& \mu_0 T \bnabla^2 \phi + \mu_0\bnabla \cdot \left( \bnabla \rho_0 V * \phi\right) (\br,t) \\
        &+ \mu_0\bnabla^2 \sqrt{\rho_0}V(\br - \br_0(t))+ \bnabla \cdot \sqrt{2T\mu_0} \bchi(\br,t).
    \end{split}
\end{equation}
In deriving Eq.~\eqref{eq:phi_rpa} from Eq.~\eqref{eq:eom_rho} we used the fact that, to linear order in $\phi/\sqrt{\rho_0}$, $\bnabla \cdot \gamma\bA n(\br,t)(\bnabla V * n)(\br,t)=0$ because $\bA$ is anti-symmetric, while the nonzero contribution from $\bnabla \cdot \gamma\bA\cdot n(\br,t) \bnabla V(\br - \br_0(t))$ is of order $\rho_0^{-1}$. The linearized equation for the density field is therefore not directly influenced by the presence of the transverse forces. Instead the latter manifestly appear in the equation of motion of the tracer. To proceed toward the weak coupling approximation, we consider the case where the interaction potential between the particle is soft, with $V(r)$ finite as $r\to 0$. Upon introducing the Fourier transform of a function $f$ by the convention $f(\bk) \equiv \int \dd \br \ee^{-i\bk\cdot\br} f(\br)$,  we obtain from Eq.~\eqref{eq:eom_tracer_wrho}:
\begin{equation}\label{eq:tracercoupled}
\begin{split}
        \dot \br_0(t) &=  \mu_0\bFext \\ &-i\mu_0(\mathbf{1} + \gamma\bA)\int \frac{\dd \bk}{(2\pi)^d}\ee^{i\bk\cdot\br_0(t)}\bk \sqrt{\rho_0} V(\bk) \phi(\bk,t) \\
        &+ \sqrt{2T\mu_0} \bxi_0(t),
\end{split}
\end{equation}
where $\gamma$ appears explicitly in the right-hand side. It is formally possible to express  $\phi(\bk,t)$ as a functional of the tracer position $\br_0$ upon integrating Eq.~\eqref{eq:phi_rpa}, and then Eq.~\eqref{eq:tracercoupled} becomes a self-contained equation for $\br_0$. We assume that the bath and the tracer have evolved for very long times before we start observing them so that the initial condition for the density fluctuation $\phi(\bk,t)$ can be overlooked, namely
\begin{equation}\label{eq:phik}
    \begin{split}
    \phi(\bk,t) &= \int_{-\infty}^t \dd s \Biggl[ -\mu_0k^2\sqrt{\rho_0}V(\bk) e^{-i\bk\cdot\br_0(s)} \\
    &+ i\sqrt{2\mu_0T}\bk\cdot\bchi(\bk,s) \Biggr]e^{-\mu_0Tk^2\left[1 + \beta \rho_0 V(\bk) \right](t-s)}, 
    \end{split}
\end{equation}
the correlations of the noise in Fourier space being $\langle\bchi(\bk,t)\otimes \bchi(\bk',t')\rangle = \mathbf{1} (2\pi)^d \delta(t-t')\delta(\bk+\bk')$. Substitution of Eq.~\eqref{eq:phik} into Eq.~\eqref{eq:tracercoupled} leads to
\begin{equation}
\begin{split}
    \dot\br_0(t) &= \mu_0 \bFext + \mu_0 \int_{-\infty}^t\dd s  \bF(\br_0(t)-\br_0(s),t-s) \\
    &+ \bXi(\br(t),t) + \sqrt{2\mu_0T}\bxi_0(t). 
\end{split}
\end{equation}
The interaction between the tracer and the bath splits into a deterministic force
\begin{equation}\label{eq:Ftracer}
    \begin{split}
    \bF(\br_0(t),t) &\equiv i\mu_0\int\frac{\dd \bk}{(2\pi)^3} (\mathbf{1}+\gamma\bA)\cdot\bk \\
    &\times k^2 \rho_0 V^2(\bk) \ee^{i\bk\cdot\br_0(t) - \mu_0Tk^2\left[1+\beta\rho_0V(\bk)\right]t} ,
    \end{split}
\end{equation}
and a Gaussian colored noise $\bXi$, independent of $\bxi_0$. The memory kernel $\langle\bXi(\br(t),t) \otimes \bXi(\br_0(t'), t')\rangle = \bG(\br_0(t)-\br_0(t'),t-t')$ has the following expression:
\begin{equation}\label{eq:Gtracer}
    \begin{split}
        \bG(\br_0(t),t) &\equiv \frac{(T\mu_0)^2}{\rho_0}\int_\bk (\mathbf{1} + \gamma\bA)\bkhat \otimes \bkhat(\mathbf{1}-\gamma\bA) \\
        &\times \frac{\left[\beta\rho_0V(\bk)\right]^2}{1 + \beta\rho_0 V(\bk)}\ee^{i\bk\cdot\br_0(t) - \mu_0Tk^2\left[1+ \beta\rho_0V(\bk) \right]t}.
    \end{split}
\end{equation}
We are now in a position to define, and then to determine, the various transport coefficients quantifying the dynamics of the tracer. Our computation rests on a path-integral formalism, which we briefly outline in the next subsection.

\subsection{Averaging over the tracer's trajectories}

We are interested in the computation of the diffusivity tensor, obtained from the velocity-velocity autocorrelation function~\cite{hargus2021odd}
\begin{equation}\label{eq:Dtensor}
    \begin{split}
    \bD &\equiv \int_0^{+\infty} \dd t \langle \dot\br_0(t)\otimes \dot\br_0(0) \rangle \\
    &= \lim_{t_f\to\infty} \langle[\br_0(t_f) - \br_0(0)]\otimes\dot \br_0(0)\rangle ,
    \end{split}
\end{equation}
with $\bFext=\mathbf{0}$, along with that of the mobility tensor $\bmu$, defined by the response to an external force as
\begin{equation}\label{eq:mutensor}
  \langle \br_0(t_f) - \br_0(0)\rangle \equiv \bmu\bFext t_f
\end{equation}
as $t_f\to\infty$ and $\bFext\to0$. The angular brackets denote averages over the realization of the noise. Using a path-integral formalism, these averages over the noise realization can be rewritten as weighted averages over the tracer's trajectories. If $f(\br_0(t_f))$ is a generic function of the tracer position at the final time $t_f$ then within the path-integral formalism we have
\begin{equation}
    \langle f(\br_0(t_f)) \rangle = \int \mathcal{D}\bp_0(t) \mathcal{D} \br_0(t)  f(\br_0(t_f)) \ee^{-S[\br_0(t),\bp_0(t)]} ,
\end{equation}
where the Janssen-De Dominicis functional integral $\int \mathcal{D}\bp_0(t) \mathcal{D} \br_0(t)$ is over all the trajectories $\br_0(t),\bp_0(t)$. The response field $\bp_0(t)$ is an auxiliary field encoding stochastic fluctuations. The trajectories are weighted by a trajectory-dependent exponential factor $\ee^{-S[\br_0(t),\bp_0(t)]}$. The corresponding action features two contributions, expressing respectively the motion of a free tracer and its interactions with the bath:
\begin{equation}
    S[\br_0(t), \bp_0(t)] = S_\text{free}[\br_0(t), \bp_0(t)] + S_\text{int}[\br_0(t),\bp_0(t)], 
\end{equation}
with
\begin{equation}
    \begin{split}
        S_\text{free}[\br_0(t),\bp_0(t)] &\equiv -i\int \dd t \bp_0(t)\cdot\left[\dot\br_0(t) -  \mu_0 \bFext \right] \\
        &+ \mu_0T \int \dd t \bp_0(t)^2  ,
    \end{split}
\end{equation}
and
\begin{equation}\label{eq:Sint}
    \begin{split}
        S_\text{int}&[\br_0(t),\bp_0(t)]\equiv i \int \dd t \int_{-\infty}^t \dd s \bp_0(t)\cdot\bF(\br_0(t) - \br_0(s),t-s) \\
        &+\int \dd t \int_{-\infty}^t \dd s \bp_0(t) \cdot \bG(\br_0(t) - \br_0(s), t-s)\cdot \bp_0(s).
    \end{split}
\end{equation}
To evaluate the dynamical averages in~\eqrefs{eq:Dtensor, eq:mutensor}, we resort to a small coupling approximation between the bath and the tracer and expand in powers of the small coupling parameter. This is carried out explicitly in the next subsection.  

\subsection{Weak-coupling approximation}

\label{sec:smallcoupling}

We now assume that the coupling between the bath and the tracer is weak with respect to the thermal forces. This is expressed by requiring that
\begin{equation}\label{eq:smallcoupling}
    h \equiv \frac{\sqrt{\rho_0} V}{T} \ll 1.
\end{equation}
From \eqrefs{eq:Ftracer, eq:Gtracer} we see that both $\bF$ and $\bG$ are of order $h^2$. From Eq.~\eqref{eq:Sint}, this implies that $S_\text{int}$ is of order $h^2$ as well. To second-order in $h$, dynamical averages can thus be computed by expanding the exponential $\ee^{-S}$:
\begin{equation}
    \langle f(\br_0(t_f))\rangle = \frac{\langle f(\br_0(t_f))\rangle_\text{free} - \langle f(\br_0(t_f))S_\text{int}\rangle_\text{free}}{1 - \langle S_\text{int} \rangle_\text{free}} + O(h^4),
\end{equation}
where $\langle\ldots\rangle_\text{free} \equiv \int \mathcal{D} \br_0(t) \mathcal{D}\bp_0(t) \ldots \ee^{-S_\text{free}}$ refers to an average over the dynamics of a free tracer. The action $S_\text{free}$ is quadratic, which ensures that linear and quadratic functionals of the trajectories $\br_0(t)$, $\bp_0(t)$ can be computed exactly. Since we are interested in the tracer's displacement, we define the one-point quantity $\Delta\br_0(t,s) \equiv \br_0(t) - \br_0(s)$ with $t\geq s$. We then have
\begin{equation}
    \begin{split}
        \langle\Delta\br_0(t,s)\rangle_\text{free} &= \mu_0 \bFext (t-s) , \\
        \langle \bp_0(t) \rangle_\text{free} &= 0, 
    \end{split}
\end{equation}
from which we recover the result that the mobility tensor for the free tracer is $\bmu_\text{free} = \mu_0 \mathbf{1}$. The two-point averages read
\begin{equation}\label{eq:twopointfree}
    \begin{split}
        \left\langle \bp_0(t) \otimes \bp_0(s) \right\rangle_\text{free} &= \mathbf{0} \\
        \left\langle \Delta\br_0(t,s) \otimes \bp_0(s') \right\rangle_\text{free} &= -i\mathbf{1}\chi_{[s,t)}(s') \\
        \langle \left[ \Delta\br_0(t,0) - \mu_0 \bFext t \right]\\ 
        \otimes \left[ \Delta\br_0(t',s') - \mu_0\bFext(t'-s')\right]\rangle_\text{free} &= 2\mathbf{1}\mu_0 T \text{L}([s,t] \cap [s',t']) ,
    \end{split}
\end{equation}
where $\chi_{[s,t)}(s')$ is the characteristic function of the interval $[s,t)$ (it is $1$ if $s' \in [s,t)$ and $0$ otherwise). The quantity $\text{L}[a,b]$ returns the length of the interval $[a,b]$. Note that for $\bFext=0$ the last equality in  Eq.~\eqref{eq:twopointfree} yields the mean-squared displacement of a free particle, $\langle \Delta \br_0(t_f,0)^2\rangle_\text{free} = 2d\mu_0Tt_f$. 

We now evaluate $\langle S_\text{int}\rangle_\text{free}$. This average requires the evaluation over the dynamical action for a free particle of two contributions. The first one is
\begin{equation}
    \begin{split}
        \langle \bp_0(t)\ee^{i\bk\cdot\Delta\br_0(t,s)}\rangle_\text{free} &= i\ee^{-\mu_0Tk^2(t-s) + i\bk\cdot\bFext(t-s)} \\
        &\times\langle \bp_0(t) \Delta\br_0(t,s) \rangle_\text{free}\cdot \bk \\
        &= \mathbf{0},
    \end{split}
\end{equation} 
where we made use of Wick's theorem and of the second equality of Eq.~\eqref{eq:twopointfree}. Similarly one can show that the other term vanishes, namely $\langle \bp_0(t)\otimes \bp_0(s) \ee^{i\bk\cdot\Delta\br_0(t,s)}\rangle=\mathbf{0}$. We thus conclude that 
\begin{equation}
    \left\langle S_\text{int} \right\rangle_\text{free} = 0.
\end{equation}
In the next two subsections we will compute the diffusion and the mobility tensors to leading order in $h$.

\subsection{Diffusion tensor}

We specialize the calculation to the case where no external force is applied to the tracer particle, $\bFext=0$. The calculation of the diffusion tensor to second order in the weak coupling $h$ requires computing two free-particle dynamical averages, stemming respectively from the drift term $\bF$ of Eq.~\eqref{eq:Ftracer} and from the memory term $\bG$ of Eq.~\eqref{eq:Gtracer}.  They are given by (with $s\leq t$):
\begin{widetext}
\begin{equation}
    \begin{split}
    \langle \Delta r_{0,a}(t_f,0) \dot r_{0,b}(0) p_{0,c}(t) \ee^{i\bk\cdot\Delta\br(t,s)} \rangle_\text{free} &= -2i\delta_{ac} \mu_0 T e^{-\mu_0Tk^2[1+\beta\rho_0V(\bk)](t-s)}k_b \chi_{[s,t)]}(0)\chi_{[0,t_f)}(t) \\
    \left\langle \Delta r_{0,a}(t_f,0)\dot r_{0,b}(0) p_{0,c}(t) p_{0,d}(s) \ee^{i\bk\cdot\Delta\br_0(t,s)}\right\rangle_\text{free} &= \left[ \delta_{bd}\delta_{ac}\delta(s-0)- 2\mu_0 T \delta_{ac}k_d k_b\chi_{[s,t)}(0)\right] \chi_{[0,t_f]}(t)\ee^{-\mu_0 T k^2(t-s)}.
    \end{split}
\end{equation}
Using these results, one can directly compute the diffusivity matrix
    \begin{equation}\label{eq:Dtracer}
        \begin{split}
            \bD(\gamma,T) &\approx \mu_0T \mathbf{1} - \lim_{t_f\to\infty} \langle \Delta \br_0(t_f,0) \otimes \br_0(0) S_\text{int}\rangle_\text{free} \\ 
            &= \mu_0T \mathbf{1} - \lim_{t_f\to+\infty}  i \int \dd t \int_{-\infty}^t\dd s \langle \Delta \br_0(t_f,0) \otimes \br_0(0)\left[\bp_0(t)\cdot \bF(\Delta\br_0(t,s),t-s))\right]\rangle_\text{free}  \\
            &- \lim_{t_f\to+\infty} \int \dd t \int_{-\infty}^t\dd s \langle \Delta \br_0(t_f,0) \otimes \br_0(0)\left[ \bp_0(t)\cdot \bG(\Delta\br_0(t,s),t-s))\cdot\bp_0(s)\right]\rangle_\text{free} \\
            &= \mu_0T \Biggl[\mathbf{1} - \frac{1}{2\rho_0}\int_\bk (\mathbf{1} + \gamma\bA)\bkhat \otimes \bkhat \frac{[\beta\rho_0 V(\bk)]^2}{\left[1+\frac{1}{2}\beta\rho_0 V(\bk)\right]^2} \\
            &+ \frac{1}{4\rho_0}\int_\bk (\mathbf{1} + \gamma\bA)\bkhat \otimes \bkhat (\mathbf{1} - \gamma\bA) \frac{[\beta\rho_0 V(\bk)]^3}{\left[1+\frac{1}{2}\beta\rho_0 V(\bk)\right]^2\left[1+\beta\rho_0 V(\bk)\right]}\Biggr].
        \end{split}
    \end{equation}
For $\gamma=0$, we obtain
\begin{equation}\label{eq:D_eq}
    \begin{split}
    \bD(0,T) &= \mu_0T\left[\mathbf{1} - \frac{1}{2\rho_0}\int_\bk \bkhat \otimes \bkhat \frac{[\beta\rho_0V(\bk)]^2}{\left[1+\frac{1}{2}\beta\rho_0 V(\bk)\right]\left[1+\beta\rho_0 V(\bk)\right]} \right] \\
    &= \mu_0T \mathbf{1} \left[ 1 - \frac{1}{2d\rho_0}\int_\bk \frac{[\beta\rho_0V(\bk)]^2}{\left[1+\frac{1}{2}\beta\rho_0 V(\bk)\right]\left[1+\beta\rho_0 V(\bk)\right]} \right],
    \end{split}
\end{equation}
where we used, in the second line, the fact that upon integration over all the wavevectors $\bkhat\otimes\bkhat$ can be replaced, using the isotropy of the integrand in Eq.~\eqref{eq:D_eq}, by $d^{-1}\mathbf{1}$. We thus recover the equilibrium result for the longitudinal diffusion constant obtained in~\cite{demery2011perturbative, demery2014generalized}. 
\end{widetext}
The equilibrium diffusion tensor of Eq.~\eqref{eq:D_eq} is diagonal, and the interaction of the tracer with the bath reduces its ability to diffuse. When $\gamma\neq 0$, the diffusion tensor acquires an antisymmetric contribution proportional to $\gamma$, and the diagonal part picks up a contribution proportional to $\gamma^2$. These additional terms lead respectively to odd diffusivity and enhanced diffusion. To see this explicitly, we can write the tensor products of Eq.~\eqref{eq:Dtracer} in the $(xyz)$ basis. In this basis, the matrix $\bA$ displayed in Eq.~\eqref{eq:defA} reads $\bA=\be_y\otimes\be_x - \be_x\otimes\be_y$. The second tensor product inside the brackets of Eq.~\eqref{eq:Dtracer} is, neglecting the term that vanishes upon integration over all the wavevectors,
\begin{equation}
    \begin{split}
        (\mathbf{1} + \gamma\bA)\bkhat \otimes \bkhat &= \sum_{i=x,y,z} \khat_i^2 \be_i\otimes\be_i \\
        &+ \gamma\left[ \be_y\otimes\be_x\khat_x^2 - \be_x\otimes\be_y\khat_y^2 \right]  .
    \end{split}
\end{equation}
The term proportional to $\gamma$ is antisymmetric upon integration over the wavevector $\bk$, and it contributes to the odd diffusivity. 

The third tensor in Eq.~\eqref{eq:Dtracer} is
\begin{equation}
    \begin{split}
    (\mathbf{1} + \gamma\bA)&\bkhat \otimes \bkhat(\mathbf{1} - \gamma\bA) = \sum_{i=x,y,z} \khat_i^2 \be_i\otimes\be_i  \\
    & + \gamma\left[ \be_x\otimes\be_y \khat_x^2 - \be_y\otimes\be_x\khat_y^2 \right] \\
    &+ \gamma\left[ \be_y\otimes\be_x \khat_x^2 - \be_x\otimes\be_y\khat_y^2 \right] \\
    &+\gamma^2\left[ \be_x\otimes\be_x \khat_x^2 + \be_y\otimes\be_y\khat_y^2\right].
    \end{split}
\end{equation}
Upon integration over $\bk$ the terms proportional to $\gamma$ cancel out, leaving only a contribution proportional to $\gamma^2$ to the longitudinal diffusion.

The diffusion tensor in the $(xyz)$ basis is therefore given by
\begin{equation}
    \begin{split}
        \bD &= D_{xx}(\gamma)\left[\be_x\otimes\be_x + \be_y\otimes\be_y\right] + D_{zz} \be_z\otimes\be_z \\
        &+ D_\perp\left(\be_y\otimes\be_x- \be_x\otimes\be_y\right) ,
    \end{split}
\end{equation}
with $D_\parallel(\gamma)$ given by
\begin{equation}\label{eq:Dparallelrpa}
    \begin{split}
        D_\parallel(\gamma)& = \mu_0T \Biggl( 1 - \frac{1}{2 d\rho_0}\int_\bk \frac{\left[\beta\rho_0V(\bk) \right]^2}{\left[1 + \beta\rho_0V(\bk)\right]\left[1 + \frac{1}{2}\beta\rho_0V(\bk)\right]} \\
        &+ \frac{\gamma^2}{4 d \rho_0}\int_\bk \frac{\left[\beta\rho_0V(\bk) \right]^3}{\left[1 + \beta\rho_0V(\bk)\right]\left[1 + \frac{1}{2}\beta\rho_0V(\bk)\right]^2}\Biggr) .
    \end{split}
\end{equation}
The odd diffusion constant reads 
\begin{equation}
    D_\perp(\gamma) =  -\gamma \frac{\mu_0 T}{2d\rho_0}\int_\bk \frac{\left[\beta\rho_0V(\bk) \right]^2}{\left[1 + \beta\rho_0V(\bk)\right]\left[1 + \frac{1}{2}\beta\rho_0V(\bk)\right]}.
\end{equation}
For systems with equilibrium dynamics (that is, when $\gamma=0$), Onsager reciprocity relations impose this quantity to vanish. However, when departing from equilibrium, as is the case in the presence of transverse forces, odd transport coefficients need not vanish as they are {\it a priori} not prohibited by time-reversal invariance. In our case, a nonzero odd diffusivity reveals the presence of a directed swirling motion for the tagged particle. Of course, for $\gamma=0$, the odd diffusivity vanishes, as expected in equilibrium.

We see from Eq.~\eqref{eq:Dparallelrpa} that the diffusivity of the tracer is enhanced by transverse forces, as the contribution proportional to $\gamma^2$ is positive. It is remarkable that this diffusion enhancement is captured even for a  system where the linearized bath relaxation is not affected by the transverse forces. An enhancement of diffusions in odd system was also found in the case where the oddity is induced by external magnetic fields~\cite{kalz2022collisions, kalz2023oscillatory}, and is not, as in our case, a consequence of the nonequilibrium dynamics of the system.  

We now take advantage of these explicit expressions to discuss the temperature dependence of the speedup. First, we observe that both terms in the integrals of Eq.~\eqref{eq:Dparallelrpa} are decreasing functions of the temperature. This implies that the quantity $D_\parallel(0)/\mu_0T$ is an increasing function of the temperature. Thus the efficiency of the transverse forces, defined as the ratio of the diffusion coefficient with and without transverse forces, is
\begin{equation}
    \frac{D_\parallel(\gamma)}{D_\parallel(0)} = 1 + \gamma^2 \frac{I}{D_\parallel(0)/\mu_0T} ,
\end{equation}
with $I\equiv  \frac{1}{2d\rho_0}\int_\bk \frac{\left[\beta\rho_0V(\bk) \right]^3}{\left[1 + \beta\rho_0V(\bk)\right]\left[1 + \frac{1}{2}\beta\rho_0V(\bk)\right]^2}$. We conclude that the efficiency of transverse forces increases upon cooling down the system. This is consistent with the increase of the odd diffusivity of the tracer. An analogous trend was recently found for binary mixture with nonreciprocal interactions between particles of different species~\cite{benois2023enhanced}.

It would be remarkable that the monotonous trend of the efficiency extends to low temperatures or high densities. In the following sections, using the mode coupling theory, we will show that this trend does not hold for very cool or very dense systems, where the small coupling approximation in Eq.~\eqref{eq:smallcoupling} breaks down. Before moving to the mode-coupling formalism, we briefly examine the mobility of the tracer which contains additional information, since the Einstein relation between the diffusivity and mobility tensors does not necessarily hold. 

\subsection{Mobility tensor}

The mobility tensor of the tracer can be obtained from the weak coupling expansion of Eq.~\eqref{eq:mutensor}:
\begin{equation}\label{eq:deltarweakcoupling}
    \langle \Delta\br_0(t_f,0) \rangle = \mu_0T\bFext t_f - \langle \Delta\br_0(t_f,0)S_\text{int}\rangle_\text{free} + o(t_f).
\end{equation}
The second term involves the following averages, which can be obtained using Wick's theorem and the dynamical averages of a free tracer displayed in Eq.~\eqref{eq:twopointfree}:
\begin{widetext}
\begin{equation}
    \begin{split}
        \langle \
        \Delta r_{0,a}(t_f,0)\bp_{0,b}(t) \ee^{i\bk\cdot\Delta\br_0(t,s)}\rangle_\text{free} 
        &= i\delta_{ab}\chi_{[0,t_f)}\ee^{-\mu_0Tk^2(t-s) + i\mu_0\bk\cdot\bFext} \\
        \langle \Delta r_{0,a}(t_f,0)p_{0,b}(t)p_{0,c}(s)\ee^{i\bk\cdot\Delta\br_0(t,s)}\rangle_\text{free}&= -i\delta_{ab} k_c\chi_{[0,t_f)}\ee^{-\mu_0Tk^2(t-s) + i\mu_0\bk\cdot\bFext}  .
    \end{split}
\end{equation}
Plugging these results into Eq.~\eqref{eq:deltarweakcoupling} we obtain 
\begin{equation}
    \begin{split}
        \langle \Delta\br(t_f,0)\rangle &\approx \mu_0 \bFext t_f - (\mathbf{1} + \gamma\bA)\bFext\frac{\mu_0}{2d\rho_0}\int_\bk \frac{\left[ \rho_0\beta V(\bk)\right]^2}{\left[1+\beta\rho_0V(\bk)\right]\left[1 + \frac{1}{2}\beta\rho_0V(\bk)\right]}t_f \\
        &\equiv \left(\mathbf{1}\mu_\parallel + \bA\mu_\perp\right)\bFext t_f .
    \end{split}
\end{equation}
\end{widetext}
When $\gamma=0$, the Einstein relation $D_\parallel(0,T) = T\mu_\parallel(0,T)$ is recovered. When $\gamma\neq0,$ the presence of a transverse component to the force on the tracer does not affect the longitudinal mobility of the tracer, and its expression is the same as in equilibrium~\cite{demery2014generalized}. However, an odd mobility coefficient~\cite{poggioli2023odd} proportional to $\gamma$ appears. If the tracer is pulled by a force $\bFext$ along a given direction in the $(xy)$ plane, it will also move along the transverse direction parallel to $\bA\bFext$: this is the physical meaning of the emergent odd mobility. As the temperature is lowered, the longitudinal mobility decreases while the odd mobility increases. 

The main take-home message of this section is the increase of the sampling efficiency upon decreasing the temperature. In what follows we analyse what happens when the small coupling approximation breaks down. When the strength of transverse forces increases, we ask about the asymptotic behavior of the efficiency. 

\section{Mode-coupling theory in the presence of transverse forces}

\label{sec:MCThardcore}

The dynamical evolution of the system of $N$ particles with positions $\br_i$ is governed by the operator $\Omega_{\gamma}$
\begin{equation}\label{eq:Omegagamma}
  \Omega_\gamma \equiv D_0 \sum_i \bnabla_i \cdot  \left[\bnabla_i - \left(\mathbf{1} + \gamma \bA \right)\beta \mathbf{F}_i \right].
\end{equation}
When $\gamma=0$, $\Omega_\gamma$ is the usual Smoluchowski evolution operator of the equilibrium dynamics. The evolution of the probability distribution $\rho(\mathbf{r}^N,t)$ of the system reads
\begin{equation}
  \p_t \rho(\mathbf{r}^N, t) = \Omega_\gamma \rho(\mathbf{r}^N,t) .
\end{equation}
We can thus write the formal expression for $\rho(\br^N,t)$ given its initial condition $\rho(\br^N)$:
\begin{equation}\label{FP}
  \rho(\br^N,t) = \ee^{\Omega_\gamma t}\rho(\br^N,0).
\end{equation}
We denote the average value of any function $f(\br^N,t)$ by $\langle f(\br^N,t)\rangle$, and it is fully determined by the knowledge of $\rho(\br^N,t)$:
\begin{equation}
  \begin{aligned}
    \left\langle f(\br^N,t) \right\rangle &\equiv \int d\br^N f(\br^N)\ee^{\Omega_\gamma t}\rho(\br^N) \\
    &= \int d\br^N \left[\ee^{\Omega^\dagger_\gamma t}f(\br^N)\right]\rho(\br^N) \\
    &= \int d\br^N \ee^{\Omega_{-\gamma} t}f(\br^N)\rho(\br^N) \\ 
    &=  \left\langle \ee^{\Omega_{-\gamma} t}f(\br^N) \right\rangle ,
  \end{aligned}
\end{equation}
where we have used the adjoint operator $\Omega^\dagger_{\gamma}$, defined as
\begin{equation}
  \Omega^\dagger_{\gamma} \equiv D_0 \sum_i  \left[\bnabla_i 
  + \left( \mathbf{1} +\gamma\bA\right)\beta \mathbf{F}_i\right]\cdot\bnabla_i ,
\end{equation}
and the following identity, which is a consequence of the breaking of detailed balance:
\begin{equation}\label{OmegaAf}
  \cdots \Omega_\gamma f(\br^N)\Biggr\rangle = \cdots \left( \Omega^\dagger_{-\gamma} f(\br^N) \right) \Biggr\rangle.
\end{equation}
The steady state solution of Eq.~\eqref{FP} is the Boltzmann distribution $\rho_\text{\tiny{B}}(\mathbf{r}^N) = \frac{\ee^{-\beta \mathcal{H}(\mathbf{r}^N)}}{\int d\mathbf{r}^N\ee^{-\beta \mathcal{H}(\mathbf{r}^N)} }$ with $\mathcal{H}(\mathbf{r}^N)\equiv \frac{1}{2}\sum_{i\neq j}V(\lvert\mathbf{r}_i-\mathbf{r}_j\rvert)$. Since we are interested in the steady state dynamics of the system, we assume that the initial condition is also sampled from the Boltzmann distribution, i.e. $\rho(\br^N)=\rho_{\text{\tiny{B}}}(\br^N)$.  

We are interested in the fluctuating density mode $n(\bq,t)$, defined as the Fourier transform of the fluctuating density field $n(\br,t)=\sum_i\delta(\br-\br_i(t)) - \rho_0$:
\begin{equation}
  n(\bq,t) \equiv \sum_i \ee^{-i\bq \cdot \br_i(t)} ,
\end{equation}
evaluated at the wavevector $\bq$, and the dynamical structure factor $S(\bq,t)$:
\begin{equation}\label{defSq}\begin{split}
  S(\bq, t) &\equiv \frac{1}{N}\left\langle n^*(\bq)\left(\ee^{\Omega^\dagger_{\gamma} t} n(\bq)\right) \right\rangle \\
  &= \frac{1}{N}\left\langle n^*(\bq)\ee^{\Omega_{-\gamma} t} n(\bq)\right\rangle.
\end{split}\end{equation}
The initial condition $S(\bq)$ is the equilibrium structure factor of the system. Another quantity of interest is the self-part of the intermediate scattering function, $F_s(q,t)$:
\begin{equation}\label{defFsq}
  F_s(\bq,t)\equiv \frac{1}{N}\sum_i \left\langle n_i^*(\bq) \ee^{\Omega_{-\gamma} t}n_i(\bq)\right\rangle ,
\end{equation}
with $n_i(\bq)\equiv \ee^{-i\bq\cdot\br_i}$. 

In the next section, we implement the Mori-Zwanzig projection operator formalism to obtain an equation of motion for the dynamical density correlations.

\subsection{Projection operator formalism}

\label{Mori-Zwanzig}

We start our calculation by introducing the Laplace transform
\begin{equation}
  S(\bq,z)\equiv \int_0^{+\infty} S(\bq,t) \ee^{-zt} ,
\end{equation}
which, once applied to the time derivative of Eq.~\eqref{defSq} yields
\begin{equation}\label{eq:Sstart}
  zS(\bq,z) - S(\bq) = \frac{1}{N}\left\langle  n^*(\bq)\Omega_{-\gamma} \frac{1}{z-\Omega_{-\gamma}}n(\bq)\right\rangle .
\end{equation}
We introduce an operator that projects along the density mode $n(\bq)$:
\begin{equation}\label{eq:P}
  \mathcal{P} \equiv \frac{1}{N S(\bq)}n(\bq)\Bigr\rangle\Bigl\langle n^*(\bq) ,
\end{equation}
and its orthogonal counterpart $\mathcal{Q}\equiv \mId-\mathcal{P}$. Using the resolvent identity~\cite{gotze2009complex}
\begin{equation}\label{eq:firstRid}
    \begin{split}
    \mQ\frac{1}{z-\Omega_{-\gamma}} &= \mQ\frac{1}{z-\Omega_{-\gamma}\mQ} \\
    &+ \mQ\frac{1}{z - \Omega_{-\gamma}\mQ}\mQ\Omega_{-\gamma}\mP\frac{1}{z-\Omega_{-\gamma}}  ,
    \end{split}
\end{equation}
we obtain 
\begin{widetext}
\begin{equation}\label{eq:Szafterresolventid}
    \begin{split}
        zS(\bq,z) - S(\bq) &=  \frac{1}{N}\left\langle  n^*(\bq)\Omega_{-\gamma} \frac{1}{z-\Omega_{-\gamma}}(\mP+\mQ)n(\bq)\right\rangle \\
        &= \Biggl[\frac{1}{NS(q)}\left\langle  n^*(\bq)\Omega_{-\gamma} n(\bq)\right\rangle + \frac{1}{NS(q)}\left\langle n^*(\bq) \Omega_{-\gamma}\mQ \frac{1}{z - \mQ\Omega_{-\gamma}\mQ}\mQ\Omega_{-\gamma}n(\bq)\right\rangle\Biggr] S(\bq,z).\\
    \end{split}
\end{equation}
Using the projection operators, we have separated the contributions to the evolution of the structure factor into an exponentially decaying part with frequency $-N^{-1}\left\langle  n^*(\bq)\Omega_{-\gamma} n(\bq)\right\rangle$ and a memory kernel. The frequency term reads
\begin{equation}
    \begin{split}
    -\frac{1}{N}\langle n(\bq)^*\Omega_{-\gamma}n(\bq)\rangle &= i\frac{D_0}{N}\bq\cdot 
    \sum_i \langle \ee^{i\bq\cdot\br_i}\cdot
    \left[ \left(\bnabla_i -\gamma \bA \beta \bF_i\right) n(\bq)\right]\rangle \\
    &= -D_0 q^2.
    \end{split}
\end{equation}
This term, which encodes the diffusive decay of the structure factor in the absence of interactions, is left unaffected by the transverse forces. 

Using an integration by parts, together with the fact that the Boltzmann distribution describes the steady state of the system, the memory kernel $\bMtilde$ can be written as
\begin{equation}
    \begin{split}
         \bq \cdot \frac{(D_0\beta)^2}{V\rho_0}\left\langle \sum_i e^{i\bq\cdot\br_i}((\mathbf{1} - \gamma\bA)\bF_i + iT\bq)\mQ \frac{1}{z-\mQ\Omega_{-\gamma}\mQ}\mQ \sum_j e^{i\bq\cdot\br_j}(\bF_i (\mathbf{1} - \gamma\bA) - iT\bq)\right\rangle\cdot\bq \equiv D_0\bq\cdot \bMtilde(\bq,z) \cdot\bq .
    \end{split}        
\end{equation}
Due to the presence of projection operators $\mQ$, the formula for $\bMtilde$ can be simplified,
    \begin{equation}
    \begin{split}
          \bMtilde(\bq,z) = \frac{(D_0\beta)^2}{V\rho_0}\left\langle \sum_i e^{i\bq\cdot\br_i}((\mathbf{1} - \gamma\bA)\bF_i + iT\bq)\mQ \frac{1}{z-\mQ\Omega_{-\gamma}\mQ}\mQ \sum_j e^{i\bq\cdot\br_j}(\bF_i (\mathbf{1} - \gamma\bA) - iT\bq)\right\rangle
         \nonumber \\
         = \frac{(D_0\beta)^2}{V\rho_0}\left\langle \sum_i e^{i\bq\cdot\br_i}((\mathbf{1} - \gamma\bA)\bF_i)\mQ \frac{1}{z-\mQ\Omega_{-\gamma}\mQ}\mQ \sum_j e^{i\bq\cdot\br_j}\bF_i (\mathbf{1} - \gamma\bA)\right\rangle .
    \end{split}        
\end{equation}
\end{widetext}
The memory kernel can be expressed in terms of the correlations between projected force density Fourier modes,
\begin{equation}\label{eq:Qj}
    \begin{split}
        \mQ\bj(\bq) &\equiv \mQ \sum_i \bF_i \ee^{-i\bq\cdot\br_i} .
    \end{split}
\end{equation}
Note that due to the projection operator we have $\ldots \mQ\sum_i \bF_i \ee^{-i\bq\cdot\br_i}\rangle = \ldots \mQ T \sum_i \bnabla_i \ee^{-i\bq\cdot\br_i}\rangle$ and it is the latter form that is used in the derivation of the mode-coupling approximation, Eq. \eqref{eq:mctvertex}. The memory kernel $\bMtilde$ thus reads
\begin{equation}\label{eq:Mtilde_as_Ktilde}
    \bMtilde(\bq,z) = (1-\gamma\bA)\bKtilde(\bq,z)(1-\gamma\bA) ,
\end{equation}
with 
\begin{equation}\label{eq:Ktilde}
    \bKtilde(\bq,z) \equiv \frac{D_0\beta^2}{V\rho_0}\left\langle \bj(\bq)^*\mQ \frac{1}{z-\mQ\Omega_{-\gamma}\mQ}\mQ \bj(\bq) \right\rangle. 
\end{equation}
Equation \eqref{eq:Szafterresolventid} becomes 
\begin{equation}\label{eq:zS_as_Mtilde}
    zS(\bq,z) - S(q) = \frac{D_0}{S(q)}\bq\cdot\left[-\mathbf{1} + \bMtilde(\bq,z)\right]\cdot\bq S(\bq,z).
\end{equation}
When $\gamma=0$, the memory kernel $\bMtilde$ is diagonal, and only the correlations among longitudinal particle currents contribute. We thus fall back onto the equilibrium case. In the presence of transverse forces, correlations arising from transverse currents contribute to the dynamics of the structure factor. Cross-correlations among longitudinal and transverse currents might as well influence the dynamics. This is the new feature entering the design of our mode-coupling approximation. In the next section, we expand on the irreducible representation of the memory kernel. 

\subsection{The irreducible memory kernel}

\label{Irreducible}

\newcommand{\Oi}{ \Omega^{\text{\tiny{irr}}}_{-\gamma} }
\newcommand{\IOmat}{ \left(\mathbf{1} - \bm\Gamma_\bq \right) }
\newcommand{\Amat}{ \left(\mathbf{1} - \gamma\bA \right) }
\newcommand{\IOmatT}{ \left(\mathbf{1} + \bm\Gamma^{\text{T}}_\bq \right)}
\newcommand{\Rirr}{ R^{\text{irr}}(z)}
  
We now introduce the irreducible memory kernel. Following Kawasaki~\cite{kawasaki1995irreducible} and Vogel and Fuchs~\cite{vogel2020stress} we define an irreducible operator
\begin{equation}\label{eq:Omegairr}
  \Omega^{\text{irr}}_{-\gamma} \equiv D_0 \mathcal{Q}\sum_j \bnabla_j \mathcal{Q}_j \cdot \left(\mathbf{1} - \gamma\bA \right) \left[ -\beta \bF_j + \bnabla_j\right] \mathcal{Q} ,
\end{equation}
with $\mathcal{Q}_j\equiv 1 - \mathcal{P}_j$ and $\mathcal{P}_j\equiv  \ee^{-i\bq\cdot\br_j}\rangle\langle \ee^{i\bq\cdot\br_j}$ a single-particle projection operator. The definition of the irreducible operator
comes from a formal extension of the equilibrium case. Using the fact that particles are statistically equivalent we obtain $\mathcal{Q}\Omega_{-\gamma} \mathcal{Q} = \Oi - \delta\Omega_{-\gamma}$, with
\begin{equation}\label{eq:deltaOmega}
    \delta\Omega_{-\gamma} = \frac{D_0\beta^2}{N} \mathcal{Q}\bj(\bq)\rangle \cdot \Amat \langle \bj(\bq)^*\mathcal{Q}.
\end{equation}
Using Eq.~\eqref{eq:deltaOmega} and the resolvent identity
\begin{equation}\label{eq:resolventid_irr}
    \begin{split}
    \frac{1}{z-\mQ\Omega_{-\gamma}\mQ} &= \frac{1}{z-\Oi} \\
    &+ \frac{1}{z-\Oi}\delta\Omega_{-\gamma}\frac{1}{z-\mQ\Omega_{-\gamma}\mQ} ,
    \end{split}
\end{equation}
we can express the reducible memory kernel $\bKtilde$ in Eq.~\eqref{eq:Ktilde} as a function of an irreducible memory kernel $\bK(\bq,z)$, where the evolution of the system is governed by $\Oi$:
\begin{equation}\label{eq:K}
    \bK(\bq,z) \equiv \frac{D_0\beta^2}{\rho_0 V}\left\langle \bj(\bq)^*\mQ \frac{1}{z-\Oi}\mQ \bj(\bq) \right\rangle. 
\end{equation}
The relationship between the reducible and the irreducible memory kernels is
\begin{equation}\label{eq:Ktilde_K}
    \bKtilde(\bq,z) = \left[\mathbf{1} +\bK(\bq,z) (\mathbf{1} - \gamma\bA)\right]^{-1}\bK(\bq,z).
\end{equation}
The introduction of the irreducible memory kernel is a necessary step in order to avoid an unphysical negative viscosity in the system, which may appear within an approximate evaluation of the reducible kernel~\cite{cichocki1987memory}.

Approximations are now needed to compute $\bK(\bq,z)$. In the next section we apply the mode-coupling approximation to the memory kernel.

\subsection{Mode-coupling expansion of the memory matrix}

\label{sec:modecoupling}

\newcommand{\nq}{n(\bq)}
\newcommand{\nk}{n(\bk)}
\newcommand{\nqk}{n(\bq - \bk)}

To perform the mode-coupling expansion, we follow Szamel and L\"owen~\cite{szamel1991mode}. The basic idea behind the mode-coupling approximation is to decompose the current field into a sum of products of density modes:
\begin{equation}\label{eq:mctexpansion}
  \mathcal{Q}\bj(\bq) \approx \frac{1}{2}\sum_{\bk}\frac{n(\bk)n(\bq - \bk)}{N^2 S(\bk)S(\bq - \bk)}\left\langle n^*(\bk)n^*(\bq-\bk)\mathcal{Q}\bj(\bq)\right\rangle ,
\end{equation}
where the factor $\frac{1}{2}$ comes from a Gaussian factorization of a static multi-point density correlator. The central mode-coupling approximation is the factorization of the time-dependent multi-point density correlator, combined to the approximation $\Oi\approx \Omega_{-\gamma}$:
\begin{equation}\label{eq:mctfactorization}
  \begin{aligned}
    &\left\langle n(\bk')^*n(\bq - \bk')^* \ee^{\Oi t}\nk\nqk\right\rangle \\
    &\approx   \left\langle n^*(\bk')e^{\Omega_{-\gamma} t}\nk\right\rangle \left\langle n^*(\bq - \bk') \ee^{\Omega_{-\gamma} t}\nqk\right\rangle \\
    &+  \left\langle n^*(\bk')\ee^{\Omega_{-\gamma} t}\nqk\right\rangle \left\langle n^*(\bq - \bk') \ee^{\Omega_{-\gamma} t}\nk\right\rangle \\
    &= N^2S(\bk,t)S(\bq - \bk,t) \left[\delta_{\bk,\bk'} + \delta_{\bk'\bq-\bk}\right] .
  \end{aligned}
\end{equation}
We now compute the expectation value in Eq.~\eqref{eq:mctexpansion} with the aid of a convolution approximation
\begin{equation}\label{eq:mctvertex}
  \begin{aligned}
    &\left\langle\nk^*\nqk^*\mathcal{Q}\bj(\bq)\right\rangle = \left\langle\nk^*\nqk^*\bj(\bq)\right\rangle \\  
    &-\frac{1}{NS(\bq)}\left\langle\nk^*\nqk^*\nq\right\rangle\left\langle\nq\bj(\bq)\right\rangle \\
    &\approx T\left\langle \nk^*\nqk\sum_i \bnabla_i \ee^{-i \bq\cdot\br_i}\right\rangle \\
    &- \frac{T}{NS(\bq)}NS(\bk)S(\bq-\bk)S(\bq)\left\langle \nq^* \sum_i \bnabla_i \ee^{-i \bq\cdot\br_i}\right\rangle\\
    &=  -T\left\langle \sum_i \bnabla_i\left[\nk^*\nqk^*\right]\ee^{-i \bq\cdot\br_i}\right\rangle \\
    &+ \frac{T}{NS(\bq)}NS(\bk)S(\bq-\bk)S(\bq)\left\langle  \sum_i \bnabla_i\left[\nq^*\right] \ee^{-i \bq\cdot\br_i}\right\rangle\\
    &= -iNT\left[ \bk S(\bq-\bk) + \left(\bq-\bk\right)S(\bk) - \bq S(\bk)S(\bq-\bk)\right] \\
    &= iNT\rho_0 S(\bk)S(\bq-\bk)\left[ \bk c(\bk) + \left(\bq-\bk\right)c(\bq-\bk)\right] ,
  \end{aligned}
\end{equation}
where in the last step we have introduced the direct correlation function $c(\bk)$, related to $S(\bk)$ via the Ornstein-Zernike relation $\rho_0c(\bk) = 1 - \frac{1}{S(\bk)}$. We can now inject the expansion \eqref{eq:mctexpansion} into the memory kernel $\bK$ of Eq.~\eqref{eq:K}. We then use Eq.~\eqref{eq:mctfactorization} and Eq.~\eqref{eq:mctvertex}. After replacing the sum over the wavevectors with an integral, $\sum_\bk \to V \int_\bk$, we obtain
\begin{equation}
    \bK(\bq,t) \approx \frac{D_0\rho_0}{2}\int_\bk \bV_{\bk,\bq}\otimes \bV_{\bk,\bq} S(\bk,t)S(\bq-\bk,t) ,
\end{equation}
where the vertex
\begin{equation}
    \bV_{\bk,\bq} \equiv \bk c(k) + (\bq-\bk) c(\lvert \bq-\bk \rvert)
\end{equation}
is the mode-coupling vertex of the equilibrium dynamics. 

To proceed further, we decompose the kernel $\bK$ over an orthonormal basis that depends on the wavevector $\bq$ and on the matrix $\bA$. The axes of this basis are defined by
\begin{equation}\label{eq:Aqbasis}
    \begin{split}
        \be_1 \equiv \be_{\bq1} &\equiv \frac{\bA^T\bA\bq}{\lvert \bA\bq \rvert} \\
        \be_2 \equiv \be_{\bq2} &\equiv \frac{\bA\bq}{\lvert \bA\bq \rvert} \\
        \be_3 \equiv \be_{\bq3} &\equiv \frac{(\mathbf{1} - \bA^T\bA)\bq}{\lvert (\mathbf{1} - \bA^T\bA)\bq\rvert}.  \\
    \end{split}
\end{equation}
Looking at the expression for $\bA$ in Eq.~\eqref{eq:defA}, we see that $\be_1$ is the normalized projection of $\bq$ on the $(xy)$ plane, $\be_2$ is the normalized vector orthogonal to $\be_1$ in the $(xy)$ plane, as selected by the matrix $\bA$, and $\be_3$ is the normalized projection of $\bq$ along the $z$-direction. We refer to this basis as the $\bA-\bq$ basis. The matrix $\bK(\bq,t)$ is diagonal in the basis defined in Eq.~\eqref{eq:Aqbasis}, within the mode-coupling approximation. To prove this, we consider the decomposition of $\bK(\bq,t)$ in the $\bA-\bq$ basis, namely 
\begin{equation}
    \bK(\bq,t) = \sum_{i,j=1}^3 K_{ij}(\bq,t)\be_i\otimes\be_j.
\end{equation}
The matrix element $K_{ij}$ reads
\begin{equation}\label{eq:Kij}
    \begin{split}
    K_{ij}(\bq,t) &= \frac{D_0\rho_0}{2}\int_\bk [k_i c(k) + (q_i - k_i)c(\lvert\bq-\bk\rvert)]\\
    &\times[k_j c(k) + (q_j - k_j)c(\lvert\bq-\bk\rvert)]S(\bk,t)S(\bq-\bk,t).
    \end{split} 
\end{equation}
We now consider the symmetries of the dynamics with transverse forces. The first symmetry  transformation we exploit is a reflection of direction $3$ (the axis $z$). This symmetry imposes 
\begin{equation}\label{eq:ref3_1} 
    \bK(\mR_3\bq,t) = \bK(\bq,t) ,
\end{equation}
with $\mR_3\bq\equiv \bq - 2\be_3(\bq\cdot\be_3)$ the operator that reflects a vector with respect to direction $3$. By construction, we have $\be_{\bq i}=\be_{\mR_3 \bq i}$ for $i=1,2$, and $\be_{\bq 3} = -\be_{\mR_3 \bq 3}$. This implies that $K_{3i}(\bq,t) = K_{3i}(\mR_3\bq,t)$ for $i=1, 2$. On the other hand, a change of variable $k_3 \to -k_3$ in the momentum integration of $K_{3i}(\mR_3\bq,t)$ shows that
\begin{equation}\label{eq:ref3_2}
    K_{3i}(\mR_3\bq,t) = -K_{3i}(\bq,t),
\end{equation}
where we used the fact that $S(\mR_3\bk,t)=S(\bk,t)$ due to the symmetries of the dynamics with transverse forces. Combining \eqrefs{eq:ref3_1, eq:ref3_2} we see that $K_{3i}(\bq,t)=-K_{3i}(\bq,t)$ for $i=1,2$, thus implying that $K_{3i}(\bq,t)=0$ for $i=1, 2$. 

The second symmetry we exploit is obtained by considering a rotation within the $(xy)$ plane by an angle $\pi/2$. The resulting invariance imposes 
\begin{equation}\label{eq:rotpi2_1} 
    \bK(\Rotpihalf\bq,t) = \bK(\bq,t) ,
\end{equation}
with $\Rotpihalf\bq\equiv \bA\bq + \be_3 q_3$ the operator that rotates a vector by an angle $\pi/2$ in the $(xy)$ plane. By construction, $\be_{\bq 3}=\be_{\Rotpihalf \bq 3}$, $\be_{\bq 1} = -\be_{\Rotpihalf \bq 2}$ and $\be_{\bq 2} = \be_{\Rotpihalf \bq 1}$. This implies $K_{12}(\bq,t) = -K_{21}(\Rotpihalf\bq,t)$. On the other hand, a change of variable $\bk \to \Rotpihalf \bk$ in the momentum integral of $K_{21}(\Rotpihalf\bq,t)$ shows that
\begin{equation}\label{eq:rotpi2_2}
    K_{21}(\Rotpihalf \bq,t) = -K_{12}(\bq,t),
\end{equation}
where we used the fact that $S(\Rotpihalf\bk,t)=S(\bk,t)$ due to the symmetries of the dynamics with transverse forces. Combining \eqrefs{eq:rotpi2_1, eq:rotpi2_2} we see that $K_{12}(\bq,t)=-K_{12}(\bq,t)$ for $i=1,2$, thus implying that $K_{12}(\bq,t)=K_{21}(\bq,t)=0$. This concludes our proof that the kernel $\bK(\bq,t)$ is diagonal in the $\bA-\bq$ basis, i.e.
\begin{equation}
    \bK(\bq,t) = \sum_{i=1}^3 K_{ii}(\bq,t)\be_i\otimes\be_i.
\end{equation}

Using this diagonal approximation, we can give an expression of $\bKtilde(\bq,z)$. Noting that the matrix $\bA$ reads, in the $\bA-\bq$ basis,
\begin{equation}
    \bA = \be_2\otimes\be_1 - \be_1\otimes\be_2,
\end{equation}
we can invert the matrix in Eq.~\eqref{eq:Ktilde_K}, substitute the result in Eq.~\eqref{eq:Mtilde_as_Ktilde} and Eq.~\eqref{eq:zS_as_Mtilde} to finally obtain
\begin{widetext}
    \begin{equation}\label{eq:zS_modecoupling}
        \begin{split}
            & zS(\bq,z)  -S(\bq) = -D_0\bq \cdot \Biggl[ \frac{1 + (1+\gamma^2)K_{22}}{(1+K_{11})(1+K_{22}) + \gamma^2 K_{11}K_{22}}\be_1\otimes\be_1 + \frac{1}{1+K_{33}}\be_3\otimes\be_3 \\
            &+ \gamma \frac{K_{11} + K_{22} + (1+\gamma^2)K_{11}K_{22}}{(1+K_{11})(1+K_{22}) + \gamma^2 K_{11}K_{22}}\left[\be_1\otimes\be_2-\be_2\otimes\be_1 \right] 
            + \frac{(1+K_{11})K_{22} + K_{11}(K_{22}-1)\gamma^2}{(1+K_{11})(1+K_{22}) + \gamma^2 K_{11}K_{22}} \be_2\otimes\be_2\Biggr]\cdot \bq S(\bq,z) ,
         \end{split}
    \end{equation}
where we have omitted the dependence on $(\bq,z)$ of the memory kernel $K_{ii}$, for simplicity. Note that only the first line of the term in the square bracket contributes to the decay of $S(\bq,z)$, due to the projection along the mode $\bq$. The off-diagonal, anti-symmetric term on the second line hints at the odd transport properties of the dynamics, which will be investigated below. 

Before turning to the discussion of longitudinal and odd transport, we study the decay of the dynamical structure factor to a (possibly nonzero) plateau. To do so, we introduce the normalized dynamical density correlations $\phi(\bq,t) \equiv \frac{S(\bq,t)}{S(\bq)}$. Its dynamics is readily obtained from Eq.~\eqref{eq:zS_modecoupling}:
\begin{equation}\label{eq:zphi-1}
            z\phi(\bq,z) - 1 = - \frac{D_0}{S(q)} \bq
            \cdot\Biggl[ \frac{1 + (1+\gamma^2)K_{22}}{(1+K_{11})(1+K_{22}) + \gamma^2 K_{11}K_{22}}\be_1\otimes\be_1 + \frac{1}{1+K_{33}}\be_3\otimes\be_3 \Biggr]\cdot\bq \phi(\bq, z).
    \end{equation}
\end{widetext}
For $\gamma=0$, we are back to the known equilibrium situation. The isotropy of the dynamics is restored, $K_{11}=K_{33}$ and we consistently find:
\begin{equation}
    z\phi(\bq,z) - 1 = -\frac{D_0}{S(q)} q^2\frac{\phi(\bq,z)}{1+K_\parallel(\bq,z)}
\end{equation}
where $K_\parallel$ is the longitudinal mode-coupling kernel
\begin{equation}
    \begin{split}
        K_\parallel(\bq,t) &= \frac{D_0\rho_0}{2q^2}\int_{\bk} \left[\bq\cdot \bk c(k) + \bq\cdot (\bq-\bk)c(\lvert\bq-\bk\rvert)\right]^2 \\
        &\times S(\bk,t)S(\bq-\bk,t) .
    \end{split}
\end{equation}
For the general situation with $\gamma\neq 0$, it is instructive to consider the case where $q_3=0$ and rewrite Eq.~\eqref{eq:zphi-1} in the time domain:
\begin{equation}
    \begin{split}
    \p_t \phi(\bq,t) + D_0q^2\phi(\bq,t) + D_0 q^2(1+\gamma^2)K_{22}*\phi(\bq,t) \\
    = -\left[\left(K_{11}  + K_{22}\right) +  (1+\gamma^2)K_{11}*K_{22}\right]*\p_t\phi(\bq,t) ,
    \end{split}
\end{equation}
where $*$ denotes a convolution in time, $f*g(t) = \int_0^t \dd t f(t-\tau)g(\tau)$. With this expression, it is clear that when $\gamma\neq 0$ the transverse currents affect both the relaxation rate and the friction kernel of the system.

In order to extract a physical picture from these equations, we inject an approximate form for the static structure factor in which the structure is sharply localized at a given wavevector. As a result, the next subsection explores the resulting so-called schematic approximation. 

\subsection{Schematic mode-coupling theory}

\label{schematic}

The first schematic approximation we consider follows the historical one~\cite{bengtzelius1984dynamics} where the structure factor is strongly peaked for a wavevector of modulus $q_0$, $S(\bq) \approx 1 + S_0\delta(\lvert\bq\rvert-q_0)$, which implies in turn that $\rho_0c(\bq)=1$ if $\lvert\bq\rvert=q_0$ and $0$ otherwise. Within this approximation, $K_{22}=0$. In this case, Eq.~\eqref{eq:zphi-1} has the following property: if $\phi(\bk,t)$ is rotationally invariant at $t=0$, then $\phi(\bk,t)$  remains rotationally invariant at all subsequent times $t$. The memory kernel thus reads
\begin{equation}
    \begin{split}
        K_{11}(q_0,t) &= K_{33}(q_0,t) \\
        &= \frac{D_0 S_0^2}{2\rho_0}\int\displaylimits_{\lvert\bk\rvert = \lvert\bq-\bk\rvert=q_0} \frac{d\mathbf{k}}{(2\pi)^3}\phi(\bk,t)\phi(\bq - \bk,t) \\
        &=  \lambda_{\text{eq}} \phi^2(q_0,t).
    \end{split}
\end{equation}
where $\lambda_{\text{eq}}\equiv\frac{\sqrt{3}}{16\pi^2}\frac{D_0S_0^2 q_0}{\rho_0}\phi^2(q_0,t)$.
Upon substitution into Eq.~\eqref{eq:zphi-1} we obtain, denoting by $\phi_2(q_0, z)$ the Laplace transform of $\phi^{2}(q_0,t)$,
\begin{equation}\label{zphi-1_eq}
  z\phi(q_0,z) - 1 =  -D_0q_0^2\frac{1}{1 + \lambda_{\text{eq}} \phi_2(q_0,z)}\phi(q_0,z).
\end{equation}
This equation is identical to the one obtained in the equilibrium case for  $\gamma=0$.

We briefly review here how an ergodicity breaking scenario is predicted from this equation~\cite{leutheusser1984dynamical}. If we denote by $f_0$ the nonergodicity parameter $f_0\equiv \lim\displaylimits_{t\rightarrow +\infty}\phi(q_0,t)$, we have
\begin{equation}\label{limz0phi}
  \begin{aligned}
    \lim\displaylimits_{z\rightarrow 0}\phi(q_0,z) &= \frac{f_0}{z} , \\
    \lim\displaylimits_{z\rightarrow 0}\phi_2(q_0,z) &= \frac{f^2_0}{z}. 
  \end{aligned}
\end{equation}
Taking the limit $z\rightarrow 0$ on both sides of Eq.~\eqref{zphi-1_eq} and keeping only the diverging part yields
\begin{equation}\label{f/1-f_eq}
  \frac{f_0}{1 - f_0} = \frac{\lambda_{\text{eq}}}{D_0q_0^2}f^2_0.
\end{equation}
Equation \eqref{f/1-f_eq} admits a nonzero solution if and only if $\frac{\lambda_{\text{eq}}}{D_0q_0^2}\geq 4$. This is the ergodicity breaking predicted by the mode-coupling theory of the colloidal glass transition in equilibrium~\cite{szamel1991mode}.

We now give an expression for the high-temperature relaxation time in the equilibrium case. Let us assume that the system is ergodic, $\lambda_{\text{eq}}<4D_0q^2$, and that $\phi(q_0,t)=e^{-\frac{t}{\tau_0}}$, from which it follows that $\phi(q_0,z=0)=\tau_0$ and $\phi_2(q_0,z=0)=\frac{\tau_0}{2}$. Equation~\eqref{zphi-1_eq} evaluated at $z=0$ gives an expression for the relaxation time $\tau_0$ 
\begin{equation}
  \tau_0 = \frac{1}{q_0^2D_0 - \frac{\lambda_{\text{eq}}}{2}}.
\end{equation}
This result is physical only for $\lambda_{\text{eq}}<2q_0^2D_0$, which corresponds to the ergodic phase. At higher values of $\lambda_{\text{eq}}$ another functional form for the decay of $\phi$ must be assumed~\cite{bengtzelius1984dynamics, leutheusser1984dynamical}. 

Implementing the conventional schematic approximation does not allow us to capture any qualitative change in the evolution equation of the dynamical structure factor. However, the fact that dynamics is accelerated is a mathematical statement, and so we should blame the nature of the schematic approximation. There is also the possibility that the mode-coupling approximation itself does not capture the acceleration provided by transverse forces, but our analysis, developed in the remainder of the paper, shows that this is not the case.  

The historical schematic approximation spuriously transforms a set of anisotropic equations for time-correlations into an isotropic one. We thus need to take a step back and consider an alternative schematic approximation that does not overlook the anisotropy, which we believe is an essential ingredient. This is what we implement in the next subsection.

\subsection{New ansatz for the ergodic phase: relaxation speedup}

\label{sec:schematic_new}

We now resort to an approximate form of the static structure factor that preserves the chiral character of the dynamics, and which, as will appear, implies a faster relaxation in the ergodic phase.

We keep the main ingredient of the schematic approximation, namely the idea that the structure factor is sharply localized for wavevectors around a given modulus $q_0$. Nevertheless, we replace the sharp delta-function with a rounded peak of nonzero width $\epsilon$, which is assumed to be small with respect to $q_0$, but nonzero. This will be enough to produce an interesting interplay of modes in various directions. In practice, we resort to the following approximate forms for $S(k)$ and for the product $S(\mathbf{k})S(\mathbf{q}-\mathbf{k})$:
\begin{equation}
  \begin{aligned}
    S(k) &\approx 1 + S_0\frac{1}{\epsilon^d}\eta\left(\frac{k-q_0}{\epsilon}\right) \\
    S(\lvert\mathbf{k}\rvert)S(\mathbf{q}-\mathbf{k}) &\approx \frac{S^2_0}{\epsilon^d}\eta\left(\frac{k-q_0}{\epsilon}\right)\eta\left(\frac{\lvert\bq-\bk\rvert-q_0}{\epsilon}\right)
  \end{aligned}
\end{equation}
with $\epsilon^{-1}\eta(k)$ a function normalized to unity with an absolute maximum at $k=0$. An example of such $\eta(k)$ could be a Gaussian function
\begin{equation}
  \epsilon^{-d}\eta(k) \equiv \frac{1}{\sqrt{2\pi\epsilon^{2d}}}e^{-\frac{k^2}{2\epsilon^2}},
\end{equation}
but the specific shape of $\eta$ is not relevant. Such a function will serve as an approximation of the Dirac distribution. Using the Ornstein-Zernike relation, the direct correlation function is given by
\begin{equation}
  \rho_0c(k) \approx \frac{\frac{S_0}{\epsilon^d}\eta\left(\frac{k-q_0}{\epsilon}\right)}{1 + \frac{S_0}{\epsilon^d}\eta\left(\frac{k-q_0}{\epsilon}\right)}.
\end{equation}
When computing $K_{22}(\bq,t)$, we focus on a wavevector $\bq$ such that $\lvert\bq\rvert=q_0$ and expand around wavevectors $\bk_0$ such that $k_0=q_0$ and $\lvert \bq - \bk_0\rvert=q_0$. We then rewrite $\bk$ as
\begin{equation}
  \bk = \bk_0 + \epsilon \bp
\end{equation}
in the integrals of Eq.~\eqref{eq:Kij}. Note that $\epsilon$ has the dimension of a wavevector, and hence $\bp$ is dimensionless. The approximate values of $k$ and $\lvert\bq-\bk\rvert$ up to order $\epsilon^2$ are then
\begin{equation}
  \begin{aligned}
      k &= \lvert \bk_0 + \epsilon\bp\rvert = q_0 + \epsilon\frac{\bp\cdot\bk_0}{q_0} + \frac{\epsilon^2}{2q_0}\left[ p^2 - \frac{\left(\bp\cdot\bk_0\right)^2}{q_0^2}\right] \\
      \lvert \bq-\bk\rvert &\approx q_0 + \epsilon \frac{\bp\cdot\left(\bq-\bk_0\right)}{q_0} + \frac{\epsilon^2}{2q_0}\left[ p^2 - \frac{\left(\bp\cdot\left(\bq-\bk_0\right)\right)^2}{q_0^2}\right].
  \end{aligned}
\end{equation}
We now expand the direct correlation function in powers of $\epsilon$. We assume  the function $\eta$ to be even, so that $\rho_0c'(q_0)=0$. We thus obtain
\begin{equation}\label{eq:expansion_c}
    \begin{split}
        \rho_0c(k) &\approx \rho_0 c(q_0) + \epsilon^2 \frac{(\bp\cdot\bk_0)^2}{2q_0^2}\rho_0c''(q_0) \\
        &= 1 + \epsilon^2 \frac{(\bp\cdot\bk_0)^2}{2S_0q_0^2}\frac{\eta''_0}{\eta_0^2} ,
    \end{split}
\end{equation}
where we used the notation $\eta_0\equiv \eta(0)$, $\eta''_0 \equiv \eta''(0)$. Substituting Eq.~\eqref{eq:expansion_c} into the expression for the memory kernel $K_{22}(\bq,t)$, given in Eq.~\eqref{eq:Kij}, we find to lowest order in $\epsilon$,
\begin{equation}
    \begin{split}
        K_{22}(\bq,t) &= \frac{D_0\rho_0}{2}\int_\bk \left(\frac{\bA\bq\cdot\bk}{|\bA\bq|}\right)^2 \left[ c(k) - c(|\bq-\bk|)\right]^2 \\
        &\times S(k)S(|\bq-\bk|) \phi(\bk,t)\phi(\bq-\bk,t) \\
        &\approx \epsilon^4 \frac{D_0(\eta''_0)^2}{8\rho_0q_0^4\eta_0^4}\int_\bp \left(\frac{\bA\bq\cdot\bk_0}{|\bA\bq|}\right)^2 (\bp\cdot\bq)^2\\
        &\times \left[ 2\bp\cdot\bk_0 - \bp\cdot\bq\right]^2 \eta\left(\frac{\bp\cdot\bk_0}{q_0}\right)\eta\left(\frac{\bp\cdot(\bq-\bk_0)}{q_0}\right) \\
        &\times \phi(\bk_0,t)\phi(\bq-\bk_0,t) .
    \end{split}
\end{equation}
The integrand is nonzero as long as $\bA\bq\neq0$. Allowing for a small momentum shell of thickness $\epsilon$ was enough for anisotropy to play a role, in contrast to the standard schematic approximation.

The previous result justifies the following approximation for the memory kernel $\bK(\bq,z)$, when $|\bq|=q_0$ and $\bA\bq \neq 0$: 
\begin{equation}
    \bK(\bq,z) = \lambda \phi_2(q_0,z)(\be_1\otimes\be_1 + \be_3\otimes\be_3) + \nu\phi_2(q_0,z)\be_2\otimes\be_2 ,
\end{equation}
with $\lambda$ and $\nu$ two parameters that depend continuously on density and temperature. Moreover, as a further simplification, we assume that all the modes relax as the ones in the $(xy)$ plane. We speculate that the anisotropy of the system would simply lead to a renormalization of $\gamma$ for our theory, thus weakly affecting the results discussed in the following.

Within the above approximation, Eq.~\eqref{eq:zphi-1} reads 
\begin{equation}\label{zphi-1_sch2}
    z\phi - 1 =  -q_0^2D_0\phi\frac{1 + \nu\phi_2(1+\gamma^2)}{1 + \phi_2(\lambda+\nu)+\phi_2^2\lambda\nu\left(1+\gamma^2\right)} .
\end{equation}
For $\gamma=0$, this expression must match its equilibrium counterpart. Therefore we must have $\lambda=\lambda_{\text{eq}}$. In the next paragraph we investigate the glass transition and the dynamic acceleration in the ergodic phase.

\subsubsection{Ergodicity Breaking}

We substitute Eq.~\eqref{limz0phi} into Eq.~\eqref{zphi-1_sch2} and keep only the leading diverging terms of order $1/z^2$:
\begin{equation}
  \lambda_{\text{eq}}\nu\left(1+\gamma^2\right)\left(1-f_0\right)f_0 = q_0^2D_0 \nu \left(1+\gamma^2\right) .
\end{equation}
This expression simplifies into Eq.~\eqref{f/1-f_eq}, namely $\frac{1}{1-f_0}=\frac{\lambda_{\text{eq}}}{q_0^2D_0}f_0$, which establishes that dynamic ergodicity breaking occurs at exactly the same location as in equilibrium.

\subsubsection{Speedup of the relaxation time}

\label{sec:schematic_tau_ergodic}

Using $\phi(q_0,z=0)=\tau_0$, $\phi_2(q_0,0)=\frac{\tau_0}{2}$ and substituting into Eq.~\eqref{zphi-1_sch2} with $z=0$ we obtain an equation for $\tau_0$:
\begin{widetext}
\begin{equation}
\tau_0^2 \left[\left(1+\gamma^2\right)\left(\frac{\lambda}{4}-q_0^2D_0\right)\nu\right] 
  + \tau_0\left[\frac{1}{2}\left(\lambda+\nu\right)-q_0^2D_0\right] + 1 = 0 .
\end{equation}
The positive solution for any value of $\gamma$ is
\begin{equation}
  \tau_0^{-1}(\gamma) = \frac{4}{-\lambda-\nu + 2q_0^2D_0 + \sqrt{\lambda^2 + \nu^2 + 4\nu q_0^2D_0(1+2\gamma^2)+4q_0^2D_0^2-2\lambda(\nu + 2\gamma^2\nu + 2q_0^2D_0)}}.
\end{equation}
\end{widetext}
This describes the relaxation time for exponential relaxation in the ergodic phase. The relaxation is exponential as long as $\lambda_{\text{eq}}<2q_0^2D_0$, as in equilibrium. As a consistency check,  $\tau_0(0)$ matches the equilibrium solution, while $\tau_0(\gamma)<\tau_0(0)$ for all values of $\gamma$. This proves that transverse forces accelerate the relaxation of the system. 

In the limit $\gamma \to \infty$, the relaxation time reads:
\begin{equation}
  \tau_0(\gamma) = \frac{1}{\gamma\nu}\sqrt{\frac{2}{\left(q_0^2D_0-\frac{\lambda}{2}\right)}}.
\end{equation}
In this limit, the relaxation time goes to $0$ linearly in $1/\gamma$. If correct, this implies a relaxation time that goes to zero for a large amplitude of the transverse forces. However, in numerical applications this method would be hindered by errors in the discretized equation of motion as increasing $\gamma$ would in practice demand smaller and smaller timesteps. Note also that the $\gamma^1$-dependence of the acceleration at large $\gamma$ differs from the naive $\gamma^2$ acceleration one might anticipate based on a simpler dimensional analysis as argued in \cite{ghimenti2023sampling}. This modified scaling results from a many-body correlation effect.

We now move to lower temperatures close to the dynamical glass transition, where the exponential form for the relaxation is no longer valid. 

\subsubsection{Early and late $\beta$-relaxations close to the transition}

We now study the time-dependence of $\phi$ near its intermediate-time plateau within the ergodic phase, but very close to the mode-coupling critical temperature. We study whether the known power-law regimes in this region, as found in \cite{bengtzelius1984dynamics}, are affected by the transverse forces. Our starting point is Eq.~\eqref{zphi-1_sch2}, which we recall here:
\begin{equation}\label{phizelb}
  \frac{q_0^2D_0 \phi(z)}{1-z\phi(z)} = \frac{1 + \left(\lambda+\nu\right)\phi_2(z) + \left(1+\gamma^2\right)\lambda\nu\phi_2(z)^2}{1 + \left(1+\gamma^2\right)\nu\phi_2(z)}.
\end{equation}
The system loses ergodicity for $\lambda \geq 4q_0^2D_0$. In the glass state close to the ergodicity breaking we set $\lambda=4q_0^2D_0(1+\eps)$ with $\eps \ll 1$, and we obtain
\begin{equation}
  \label{plateauweakglass}
  \lim_{t\to+\infty}\phi(t) = \frac{1}{2}\left( 1 + \eps^{1/2}\right).
\end{equation}
If instead we set
\begin{equation}
  \lambda = 4q_0^2D_0\left(1-\eps \right)
\end{equation}
we are in the ergodic phase close to the glass transition. To describe the approach of $\phi$ to the plateau we assume, in the limit $\eps \to 0$ the following scaling form, taken from Eq.~\eqref{plateauweakglass}:
\begin{equation}\label{eq:phi_close_to_plateau}
  \phi(t) = \frac{1}{2} + \eps^{1/2}g(\tau) ,
\end{equation}
with $\tau \equiv \eps^\alpha t$ a rescaled time. We begin by assuming that $\alpha>1$ and we shall later check that this is self-consistently correct. When considering the Laplace transform, we use a rescaled variable $z \equiv \eps^{\alpha}\zeta$. The Laplace transforms of $\phi(t)$ and $\phi^2(t)$ are therefore
\begin{equation}\label{phizeta}
  \begin{aligned}
    \phi(\zeta) &= \eps^{-\alpha}\left[\frac{1}{2\zeta} + \eps^{1/2}g(\zeta)\right]  , \\
    \phi_2(\zeta) &= \eps^{-\alpha}\left[\frac{1}{4\zeta} + \eps^{1/2}g(\zeta) + \eps g_2(\zeta)\right].
  \end{aligned}
\end{equation}
We thus substitute Eq.~\eqref{phizeta} into Eq.~\eqref{phizelb} and search for the leading order in $\eps$. To do so, we assume that we are in a regime where $\lvert \eps^{1/2}g(\tau)\rvert\ll 1$. The orders $\eps^0$ and $\eps^{1/2}$ are self-consistently satisfied, while at order $\eps$ we obtain
\begin{equation}
  \label{gepsilon}
  8\zeta g^2(\zeta) - 4g_2(\zeta) = -\frac{1}{\zeta}.
\end{equation}
This is the same equation as that obtained in equilibrium~\cite{bengtzelius1984dynamics}. The exponents controlling the approach to and the departure from the plateau are left unchanged by the transverse dynamics.

For completeness, we also recall how to compute these exponents below. For the early $\beta$-relaxation we take
\begin{equation}
  g(\tau) \equiv a_0\tau^{-a}
\end{equation}
for $\tau \ll 1$, which implies that
\begin{equation}
  \begin{aligned}
    g(\zeta) &= a_0\zeta^{a-1}\Gamma(1-a) , \\
    g_2(\zeta) &= a_0^2\zeta^{2a-1}\Gamma(1-2a),
  \end{aligned}
\end{equation}
for $\zeta\gg 1$, with $\Gamma(x)\equiv \int_0^{+\infty} t^{x-1}e^{-t}dt$ the gamma function. Substitution into Eq.~\eqref{gepsilon} yields to leading order in $\zeta$ 
\begin{equation}
  4a_0^2\zeta^{2a-1}\left[2\Gamma(1-a)^2-\Gamma(1-2a)\right]=0.
\end{equation}
This equation has two solutions $a=-1$ and $a=0.395$. Among these, we retain only the physical one with $a=0.395$. This solution is consistent with the earlier assumption that $\alpha>1$. In fact, if we want
\begin{equation}
  \phi(t) = \frac{1}{2} + a_0t^{-a}
\end{equation}
to hold, then we must have $\alpha=1/2a>1$. We also assess in which range of times this solution is valid. Since we want $\lvert \eps^{1/2}g(\tau)\rvert\ll 1$ and $\tau\ll1$ we must have
\begin{equation}
  \eps^{\frac{1}{2a}}a_0^{1/a}\ll \tau \ll 1 ,
 \end{equation}
which implies
\begin{equation}
  a_0^{1/a} \ll t \ll \eps^{-1/2a}.
\end{equation}

For the late $\beta$-relaxation we assume
\begin{equation}
  g(\tau) \equiv -b_0\tau^{b}
\end{equation}
for $\tau \gg 1$, which implies that
\begin{equation}
  \begin{aligned}
    g(\zeta) &= -b_0\zeta^{-b-1}\Gamma(1+b) , \\
    g_2(\zeta) &= b_0^2\zeta^{-2b-1}\Gamma(1+2b), 
  \end{aligned}
\end{equation}
for $\zeta\ll 1$. Substitution in Eq.~\eqref{gepsilon} yields to leading order in $\zeta$ 
\begin{equation}
  4a_0^2\zeta^{-2b-1}\left[2\Gamma(1+b)^2-\Gamma(1+2b)\right]=0.
\end{equation}
This equation has the solution $b=1$. We also assess in which range of times this solution is valid. Since we want $\lvert \eps^{1/2}g(\tau)\rvert\ll 1$ and $\tau\gg 1$ we must have
\begin{equation}
  1\ll \tau \ll \eps^{\frac{1}{2b}}b_0^{-1/b} ,
\end{equation}
which implies
\begin{equation}
  \eps^{-1/2a} \ll t \ll \eps^{-\frac{1}{2a}-\frac{1}{2b}}b_0^{-1/b}.
\end{equation}

This approach does not allow us to compute the values of the two constants $a_0$ and $b_0$. These will however exhibit a dependence on $\gamma$, which, we expect, should result in a marginal shortening of the time required to approach and leave the plateau in the temperature regime close to the ergodicity breaking transition.

\subsubsection{Divergence of the relaxation time}

In this subsection we show that the exponent characterizing the divergence of the relaxation time close to criticality is also unchanged by the transverse forces. 

We start by assuming the following form for $\phi(t)$ for $t\gg a_0^{1/\alpha}$:
\begin{equation}
    \phi(t) \equiv \frac{1}{2} e^{2\eps^{1/2}g(\eps^\alpha t)} .
\end{equation}
This expression naturally gives Eq.~\eqref{eq:phi_close_to_plateau} whenever $\lvert \eps^{1/2}g(\eps^\alpha t)\rvert \ll 1$, which corresponds to the plateau regime investigated earlier. When $t\gg \eps^{-\frac{1}{2a}-\frac{1}{2b}}b_0^{-1/b}$ we have
\begin{equation}
  \label{eq:phi_taualpharegime}
    \phi(t) = \frac{1}{2}e^{-2b_0\eps^{\alpha+1/2}t}.
\end{equation}
The Laplace transforms read
\begin{equation}\label{eq:div_tau_laplacetransform}
    \begin{split}
        \phi(z) &= \frac{1}{2(z + 2b_0\eps^{\alpha+1/2})} , \\
        \phi_2(z) &= \frac{1}{4(z + 4b_0\eps^{\alpha+1/2})} .
    \end{split}
\end{equation}
Using the scaling $z=\eps^{\alpha+1/2} \zeta$ we see that, to leading order in $\eps$, Eq.~\eqref{eq:div_tau_laplacetransform} satisfies Eq.~\eqref{phizelb}, which confirms that the ansatz of Eq.~\eqref{eq:phi_taualpharegime} is correct. Our schematic mode-coupling theory therefore predicts that, close to the critical temperature $T_\text{MCT}$, the relaxation time $\tau_\alpha$ diverges as a power law, 
\begin{equation}
    \tau_\alpha \sim (T-T_\text{MCT})^{-1/2a + 1/2b}, 
\end{equation}
with $\frac{1}{2a} + \frac{1}{2b} \approx 1.7658$, which is again the same exponent as in equilibrium~\cite{bengtzelius1984dynamics}.

\subsection{Beyond the schematic approximation}

Using insights gained from the schematic approximation, we now prove some general properties of the dynamics with transverse forces, as derived in Eq.~\eqref{eq:zphi-1}, in particular regarding the location of the glass transition and the acceleration of the dynamics.

\subsubsection{Location of the glass transition}

To locate the glass transition, we assume that when ergodicity breaking occurs, the plateau is the same for all wavevectors with the same modulus:
\begin{equation}\label{eq:plateaus}
    \begin{split}
    \phi(\bq,z\to0) &\approx \frac{\phi_\infty(\bq)}{z} , \\
    K_{11}(\bq,z\to0) &\approx K_{33}(\bq,z\to0) \approx \frac{K_{\parallel,\infty}(\bq)}{z} , \\
    K_{22}(\bq,z\to0) &\approx \frac{K_{\perp,\infty}(\bq, z\to 0)}{z}.
    \end{split}
\end{equation}
Taking the limit $z\to0$ in Eq.~\eqref{eq:zphi-1}, and using Eq.~\eqref{eq:plateaus}, the contribution from transverse forces cancels, leading to 
\begin{equation}
    q^2D_0 \frac{\phi_{\infty}(\bq)}{1-\phi_{\infty}(\bq)} = K_{\parallel,\infty}(\bq).
\end{equation}
This equation is the same as the one governing the equilibrium case. This establishes that the glass transition takes place for the same value of the parameters (density, temperature) as in equilibrium when $\gamma=0$. 

\subsubsection{Acceleration in the ergodic phase}

In the ergodic phase, however, some acceleration can be achieved even if the location of the glass transition is the same. To support this assertion, we begin by self-consistently assuming that transverse forces accelerate the decay of the dynamical structure factor. In practice, we start from the postulate that 
\begin{equation}
    \phif(\bq) \leq \phi_\text{\tiny{eq, f}}(\bq) ,
\end{equation}
where $\phif \equiv \phi(\bq,z=0) \equiv \int_0^{+\infty} \dd\tau \phi(\bq,\tau)$ is the time-integrated normalized dynamical structure factor, and $\phi_\text{\tiny{eq,f}}$ is the same quantity for equilibrium dynamics, obtained from Eq.~\eqref{eq:zphi-1} when $\gamma=0$. Due to the mode-coupling expression of the memory kernel, Eq.~\eqref{eq:Kij}, it follows that $K_{ij,\text{\tiny{f}}}(\bq) \equiv K_{ij}(\bq,z=0)\leq K_{ij,\text{\tiny{eq,f}}}(\bq) \equiv K_{ij,\text{\tiny{eq}}}(\bq,z=0)$. This means that 
\begin{equation}
    \frac{1 + (1+\gamma^2)K_{22,\text{\tiny{f}}}}{(1+K_{11,\text{\tiny{f}}})(1+K_{22,\text{\tiny{f}}}) + \gamma^2 K_{11,\text{\tiny{f}}}K_{22,\text{\tiny{f}}}}\geq \frac{1}{1+K_{11,\text{\tiny{eq,f}}}}, 
\end{equation}
where we omitted the dependence on the $\bq$ argument for clarity. This inequality  self-consistently proves that the transverse force dynamics accelerates the decay rate of $\phi$ in the ergodic phase. 

\subsubsection{Limiting behavior for large $\gamma$}

We explore the asymptotic behavior of the relaxation in the ergodic in the presence of very strong transverse forces. In the ergodic phase, the $z\to0$ limit of Eq.~\eqref{eq:zphi-1} reads
\begin{equation}\label{eq:ergodicregion}
    \begin{split}
        1 &= - D_0 \bq
        \cdot\Biggl[ \frac{1 + (1+\gamma^2)K_{22,\text{f}}}{(1+K_{11,\text{f}})(1+K_{22,\text{f}}) + \gamma^2 K_{11,\text{f}}K_{22,\text{f}}} \be_1\otimes\be_1 \\
        &+ \frac{1}{1+K_{33,\text{f}}}\be_3\otimes\be_3 \Biggr]\cdot\bq \phi_{{\text{f}}} .
    \end{split}
\end{equation}
The analysis of the exponential relaxation in the schematic approximation suggests that, in the limit $\gamma\to \infty$, one should expect
\begin{equation}\label{eq:philargegamma}
    \lim_{\gamma\to\infty} \phi_\text{f} = \frac{\overline{\phi_\text{f}}}{\gamma},
\end{equation}
where $\overline{\phif}$ is a constant that does not depend on $\gamma$. Our starting point is the assumption that a similar scaling holds for the memory kernels, namely
\begin{equation}\label{eq:Klargegamma}
    \lim_{\gamma\to\infty} K_{ii,\text{f}} =\frac{\overline{K}_{ii,\text{f}}}{\gamma},
\end{equation}
with $\overline{K}_{ii,\text{f}}$ a quantity independent from $\gamma$. This scaling can be explicitly checked in the case of exponential relaxation. If we substitute the asymptotic behaviors of \eqrefs{eq:philargegamma, eq:Klargegamma} into Eq.~\eqref{eq:ergodicregion}, the dependence on $\gamma$ disappears. This result consistently demonstrates that the relaxation time decreases linearly in $\gamma^{-1}$ in the limit of strong transverse forces. Note that, by contrast to the discussion in Sec.~\ref{sec:schematic_tau_ergodic} which was limited to exponential decays, this result is valid over the whole ergodic phase. 

In this section, we have investigated the relaxation properties of the collective density modes in the presence of transverse forces, demonstrating the existence of a speedup, and establishing its asymptotic behavior for strong drift. We have also realized that transverse forces do not affect the location of the transition to the nonergodic regime. This, in turn, triggers a number of questions related to the microscopic mechanisms and dynamical pathways opened by transverse forces that support  these observed collective behaviors. To answer this question, we turn to the analysis of the motion of an individual particle. The transport coefficients of this tracer, along with the diffusivity and mobility tensors, will help us shape a qualitative picture.

\section{Dynamics of a tracer with transverse forces in the mode-coupling approach}

\label{sec:MCTtracer}

We now analyze the motion of a tracer within the mode-coupling framework. We first introduce the general setting and the relevant dynamical quantities. 

Following Ernst and Dorfman~\cite{ernst1975nonanalytic}, we consider the motion of a tagged tracer particle with label $0$ in a liquid of $N$ particles with transverse forces. The tracer initial position at time $t=0$, $\br_0$, is fixed at the origin, and it influences the distribution of the surrounding bath. The initial condition for the probability distribution of the total system thus reads
\begin{equation}\label{eq:Ptracertzero}
    P_{N+1}(\br^{N+1},t=0) = V\delta(\br_0)\rho_\text{\tiny B}(\br^{N+1}).
\end{equation}
Averages over the initial distribution $P_{N+1}(\br^{N+1},0)$ will be denoted by $\langle\ldots\rangle_0$. Averages over the Boltzmann distribution are denoted by $\langle\ldots\rangle$, as usual. Starting from $t=0$,  a constant external force $\bF_\text{ext}$ is applied on the tracer. The evolution operator becomes $\Omega^\text{ext}_\gamma$, defined as
\begin{equation}
    \begin{split}
        \Omegat =& \Omega_\gamma + \delta\Omega^{\text{ext}} \\
        =&  D_0 \sum_i \bnabla_i \cdot  \left[\bnabla_i - \left(\mathbf{1} + \gamma \bm A \right)\beta \mathbf{F}_i\right] \\
        &- D_0\bnabla_0\cdot\beta\bF_\text{ext}. 
    \end{split}
\end{equation}
We denote the Fourier transform of the tracer density as
\begin{equation}\label{eq:tracerdensity}
    \nt(\bq) \equiv \ee^{-i\bq\cdot\br_0}.
\end{equation} 
The large-time, long-wavelength limit of this Fourier transform can be used to obtain relevant transport coefficients. A small $\bq$ expansion of the time derivative of Eq.~\eqref{eq:tracerdensity} yields constitutive equations of motion for the tracer
\begin{equation}
    \begin{split}
        \langle\p_t \nt(\bq,t)\rangle_0 &= -i\bq\cdot\langle \dot\br_0(t) \rangle_0 \\
        &- \bq\cdot\int_0^t\dd\tau \langle\dot\br_0(\tau)\otimes\dot\br_0(0) \rangle_0\cdot\bq \\
        &- \bq\cdot \langle \dot\br_0(t) \otimes \br_0 \rangle_0\cdot\bq\\
        &+ O(q^3) \\
        &\approx -i\bq\cdot \bmu\cdot \bFext - \bq\cdot\bD\cdot\bq .
    \end{split}
\end{equation}
In the second line, we have taken a large time limit, $t\to\infty$, and used the definition of the diffusivity tensor $\bD$, given by Eq.~\eqref{eq:Dtensor}, and the mobility tensor $\bmu$, given by Eq.~\eqref{eq:mutensor}. Note that the mobility and the diffusivity are computed within a linear response formalism, where the intensity of the external force is  small, $\lvert \bFext\rvert \to0$. Moreover, as for the weak coupling analysis of the tracer, we are interested in the expression of the diffusivity tensor obtained for $\bFext=0$, and neglect the dressing of the diffusivity that comes from the presence of the dragging external force. 

\subsection{Equation of motion of the tracer}

We introduce a projection operator tailored to the space of the tracer density fluctuations
\begin{equation}
    \mP_0 \equiv \sum_\bq \ldots\nt^*(\bq)\rangle\langle\nt(\bq)\ldots
\end{equation}
and the associated orthogonal projector $\mQ_0 \equiv \mId - \mP_0$. Note that, in contrast with the other projection operators used in this work, $\mP_0$ contains a summation over all wavevectors, since these are all included inside the $\delta$-function of the initial condition in Eq. \eqref{eq:Ptracertzero}. With this definition, we have 
\begin{equation}
    \begin{split}
        \mP_0 P_{N+1}(\br^{N+1};t=0) &= \frac{1}{V}\sum_\bq \nt(\bq)\rangle\langle \nt^*(\bq)V\delta(\br_0)\rangle \\
        &= \sum_\bq \nt(\bq)\rangle\\
        &= V\delta(\br_0)\rho_\text{\tiny{B}}(\br^{N+1}) \\
        &= P_{N+1}(\br^N;t=0) ,
    \end{split}
\end{equation}
and thus 
\begin{equation}
    \mQ_0 P_{N+1}(\br^N,t=0) = 0.
\end{equation}
Our goal is to derive an equation of motion for the evolution of the average tracer density, $\langle n_0(\bq)\ee^{\Omegat t}\rangle_0 $. In this notation, the probability distribution $P_{N+1}(\br^{N+1},0)$ stands to the right of the evolution operator $\ee^{\Omegat t}$, which acts on the said distribution. The time derivative of this quantity reads, after a Fourier transformation
\begin{widetext}
\begin{equation}
    \begin{split}
        \left\langle \nt(\bq) \Omegat \frac{1}{z-\Omegat}\right\rangle_0 &= \left\langle \nt(\bq) \Omegat \mP_0 \frac{1}{z-\Omegat}\right\rangle_0 + \left\langle \nt(\bq) \Omegat \mQ_0 \frac{1}{z-\Omegat}\right \rangle_0 \\
        &= \Biggl[\langle \nt(\bq) \Omegat \nt^*(\bq)\rangle + \left\langle \nt(\bq)\Omegat \mQ_0 \frac{1}{z-\Omegat \mQ_0}\mQ_0\Omegat \nt^*(\bq)\right\rangle\Biggr] \nt(\bq,z),
    \end{split}
\end{equation}
with $\nt(\bq,z) = \left\langle \nt(\bq) \frac{1}{z-\Omegat}\right\rangle_0$. We used the same resolvent identity as in Eq.~\eqref{eq:resolventid_irr} (with $\mP_0$ and $\mQ_0$ instead of $\mP$ and $\mQ$), namely
\begin{equation}
    \frac{1}{z-\Omegat} = \frac{1}{z-\Omegat \mQ_0} 
    + \frac{1}{z-\Omegat \mQ_0}\mQ_0\Omegat\mP_0\frac{1}{z-\Omegat} ,
\end{equation}
together with the fact that $\left\langle \ldots \mQ_0 \frac{1}{z-\Omegat \mQ_0}\right\rangle_0=0$. The frequency matrix reads
\begin{equation}
    \langle \nt(\bq)\Omegat\nt^*(\bq)\rangle = -D_0q^2 - iD_0\beta\bq\cdot\bFext,
\end{equation}
and it contains the mobility of a free tracer, $D_0\beta$. 

The memory kernel reads instead 
    \begin{equation}\label{eq:memorykerneltracerfirststep}
        \begin{split}
            &\left\langle \nt(\bq) \Omegat \mQ_0 \frac{1}{z-\mQ_0\Omegat \mQ_0}\mQ_0\Omegat\nt^*(\bq)\right\rangle =  \left\langle \nt^*(\bq) \Omega_{-\gamma,\text{\tiny{ext}}} \mQ_0 \frac{1}{z-\mQ_0\Omega_{-\gamma,\text{\tiny{ext}}} \mQ_0}\mQ_0\Omega_{-\gamma,\text{\tiny{ext}}}\nt(\bq)\right\rangle \\
            &= -i\bq\cdot D_0^2 \Biggl\langle \ee^{i\bq\cdot\br_0}\left[ (\mathbf{1} - \gamma\bA)\cdot\beta\bF_0-i\bq + \beta\bFext\right]\mQ_0 \frac{1}{z-\mQ_0\Omega_{-\gamma,\text{\tiny{ext}}}\mQ_0}\mQ_0\bnabla_0\ee^{-i\bq\cdot\br_0}\Biggr\rangle\cdot\left[(1 - \gamma\bA)\cdot i\bq - \beta\bFext \right].
        \end{split}
    \end{equation}
\end{widetext}
The first term $\bFext$ on the right hand side does not contribute, since it belongs to the space of the tracer's density modes. It  thus vanishes when the projector $\mQ_0$ acts on its right. To extract the mobility of the tracer, we are interested in terms linear in $\bFext$. One of these terms comes from the last $\bFext$ appearing on the right-hand side of Eq.~\eqref{eq:memorykerneltracerfirststep}. Another term comes in principle from the expansion of the operator $\Omegat$. However, the physical meaning of this term is a dressing of the diffusion matrix by means of the external force, as it yields a contribution proportional to $q^2$. The tracer's equation of motion can thus be cast in the form
\begin{equation}
    \begin{split}
        z\langle& \nt(q,z)\rangle_0 - 1 = [ -i\bq\cdot\bmu(\bq,z)\cdot\bFext \\
        &- \bq\cdot\left[\bD(\bq,z)+|\bFext|\delta\bD(\bq,z)\right]\cdot\bq ]\bn(\bq,z)   ,
    \end{split}
\end{equation}
where $\delta\bD(\bq,z)$ is the correction to the diffusivity tensor due to the applied external force. We are not interested in this term, and we focus on the mobility and diffusivity tensors, that read respectively
\begin{equation}\label{eq:mobility_tensor_projection}
    \bmu(\bq,z) = D_0\beta \left[\mathbf{1} - (\mathbf{1}-\gamma\bA)\bKtilde_0(\bq,z) \right] ,
\end{equation}
and 
\begin{equation}\label{eq:diffusivity_tensor_projection}
    \bD(\bq,z) = D_0\left[\mathbf{1} - (\mathbf{1}-\gamma\bA)\bKtilde_0(\bq,z)(\mathbf{1} - \gamma\bA)\right].
\end{equation}
The tracer-memory kernel can be expressed in terms of the projected longitudinal tracer force density Fourier modes
\begin{equation}\label{eq:Ktilde0}
    \bKtilde_0(\bq,z) \equiv 
    D_0\beta^2 \left\langle\bj_0^*(\bq)\mQ_0\frac{1}{z-\mQ_0\Omega_{-\gamma}\mQ_0}\mQ_0 \bj_0(\bq)\right\rangle ,
\end{equation}
with $\mQ_0 \bj_0(\bq)\equiv \mQ_0 \bF_0 \ee^{-i\bq\cdot\br_0}$. Note that due to the projection operator we have $\ldots \mQ_0 \bF_0 \ee^{-i\bq\cdot\br_0}\rangle = \ldots \mQ_0 T \bnabla_0 \ee^{-i\bq\cdot\br_0}\rangle$ and it is the latter form that is used in the definition of the irreducible evolution operator, Eq. \eqref{eq:K0} and the derivation of the mode-coupling vertex, Eq. \eqref{eq:tmctvertex}.

Equations (\ref{eq:mobility_tensor_projection}, \ref{eq:diffusivity_tensor_projection}) express the transport coefficients of the tracer as a function of the current-current, memory tensor $\bKtilde_0$. As usual in a mode-coupling approach, we first give an irreducible representation of this memory kernel and then expand it within the mode-coupling approximation. 

The reduction of $\bKtilde$ is achieved by introducing an irreducible evolution operator
\begin{equation}
    \Omegairr = \Omega_{-\gamma} + D_0\beta^2\sum_\bk \mQ_0 \bj_0(\bk)\rangle\cdot (\mathbf{1} - \gamma\bA)\cdot\langle\bj_0^*(\bk) \mQ_0
\end{equation}
and exploiting the identity
\begin{equation}
    \begin{split}
        \frac{1}{z-\mQ_0\Omega_{-\gamma}\mQ_0} &= \frac{1}{z-\Omegairr} \Biggl[1-\frac{D_0\beta^2}{V}\sum_\bk \mQ_0 \bj_0(\bk)\rangle\\
        &\cdot (\mathbf{1} - \gamma\bA)\cdot\langle\bj_0^*(\bk) \mQ_0 \frac{1}{z-\mQ_0\Omega_{-\gamma}\mQ_0}\Biggr].
    \end{split}
\end{equation}
We obtain from Eq.~\eqref{eq:Ktilde0}
\begin{equation}
   \bKtilde_0(\bq,z) = \left[\mathbf{1} + \bK_0(\bq,z)(\mathbf{1} - \gamma\bA) \right]^{-1}\cdot\bK_0(\bq,z),
\end{equation}
where we have introduced the irreducible memory kernel for the tracer
\begin{equation}\label{eq:K0}
    \bK_0(\bq,z) \equiv \beta^2 D_0 \left\langle \bj^*_0(\bq)\mQ_0 \frac{1}{z-\Omegairr}\mQ_0 \bj_0(\bq) \right\rangle.
\end{equation}
As a consequence of the linear response treatment in the definition of the $\bK_0(\bq,z)$ the evolution operator involves the dynamics with transverse forces in the absence of the external perturbation. This allows us to simplify the tracer's memory kernel, as was done previously for the memory kernel related to the dynamics structure factor, $\bK(\bq,z)$. Before discussing further this point, we apply the mode-coupling approximation scheme to $\bK_0$.

\subsection{Mode-coupling expansion of the tracer's memory kernel}

The mode-coupling expansion can be carried out in an analogous way to what was done in Sec. \ref{sec:modecoupling}. The tracer's current is expanded along the following density product:
\begin{equation}\label{eq:j0_mct_exp}
  \mathcal{Q}_0\bj_0(\bq) \approx \sum_{\bk}\frac{n(\bk)n_0(\bq - \bk)}{N S(\bk)}\left\langle n^*(\bk)n_0^*(\bq-\bk)\mathcal{Q}_0\bj_0(\bq)\right\rangle .
\end{equation}
The Gaussian approximation now becomes
\begin{equation}
  \begin{aligned}
    &\left\langle n^*(\bk')n_0^*(\bq - \bk') \ee^{\Omegairr t}\nk n_0(\bq-\bk)\right\rangle \\ 
    &\approx   \left\langle n^*(\bk')\ee^{\Omega_{-\gamma} t}\nk\right\rangle \left\langle n_0^*(\bq - \bk') \ee^{\Omega_{-\gamma} t}n_0(\bq-\bk)\right\rangle \\
    &+  \left\langle n^*(\bk')\ee^{\Omega_{-\gamma} t}n_0(\bq-\bk)\right\rangle \left\langle n_0^*(\bq - \bk') \ee^{\Omega_{-\gamma} t}\nk\right\rangle \\
    &= N S(\bk,t)F_s(\bq - \bk,t)\delta_{\bk,\bk'} + S(\bk,t)S(\bq-\bk,t)\delta_{\bk',\bq-\bk} \\
    &\approx N S(\bk,t)F_s(\bq - \bk,t)\delta_{\bk,\bk'} .
  \end{aligned}
\end{equation}
In the last passage we have neglected the term of order $1$ compared to the term of order $N^2$. 
In addition, we noted that since the tracer is equivalent to any other particle, we have $\left\langle n_0^*(\bq - \bk') \ee^{\Omega_{-\gamma} t}n_0(\bq-\bk)\right\rangle = F_s(\bq-\bk)$, with the self-intermediate correlation function $F_s$ defined in Eq. \eqref{defFsq}.

We now compute the average in Eq. \eqref{eq:j0_mct_exp} with the help of a convolution approximation
\begin{equation}\label{eq:tmctvertex}
  \begin{aligned}
    &\left\langle\nk^*n_0^*(\bq-\bk)\mathcal{Q}_0\bj_0(\bq)\right\rangle \\ 
    &= \left\langle\nk^*n_0^*(\bq-\bk)\bj_0(\bq)\right\rangle \\
    &-\left\langle \nk^*n_0^*(\bq-\bk)n_0(\bq)\right\rangle\left\langle n^*_0(\bq)\bj_0(\bq)\right\rangle \\
    &= T\left\langle \nk^*n_0^*(\bq-\bk)\bnabla_0 \ee^{-i \bq\cdot\br_0}\right\rangle \\
    &- TS(k)\left\langle n^*_0(\bq) \bnabla_0 \ee^{-i \bq\cdot\br_0}\right\rangle\\
    &=  -T\left\langle \bnabla_0\left[\nk^*n_0^*(\bq-\bk)\right]\ee^{-i \bq\cdot\br_0}\right\rangle \\
    &+ TS(k)\left\langle  \bnabla_0\left[n^*_0(\bq)\right] \ee^{-i \bq\cdot\br_0}\right\rangle\\
    &= -iT\left[ \bk + \left(\bq-\bk\right)S(k) - \bq S(k)\right] \\
    &= -iT\rho_0 S(k)c(k)\bk ,
  \end{aligned}
\end{equation}
with $\rho_0c(k)= 1 - \frac{1}{S(k)}$ the direct correlation function. With these results, substitution of Eq.~\eqref{eq:j0_mct_exp} in the irreducible memory kernel of Eq. \eqref{eq:K0} yields
\begin{equation}
    \bK_0(\bq,t) \approx D_0\rho_0\int_\bk \bk\otimes\bk c(k)^2F_s(\bq-\bk,t)S(\bk,t). 
\end{equation}
This is the mode-coupling expression of the tracer's memory kernel. Using the symmetries of the dynamics, as done in Sec.~\ref{sec:modecoupling}, one can show that $\bK_0$ is diagonal in the $\bA-\bq$ basis given by Eq.~\eqref{eq:Aqbasis}:
\begin{equation}\label{eq:K0_mct}
    \bK_0(\bq,t) = \sum_i K_{0,ii}(\bq,t)\be_i\otimes\be_i.
\end{equation}
Now that we have a diagonal decomposition of the tracer's memory kernel, we can focus on physical quantities of interest for the tracer dynamics. 

\subsection{Self-intermediate scattering function}

The dynamics of the self-intermediate scattering function $F_s$ can be read from the tracer's dynamics. It is given by
\begin{widetext}
    \begin{equation}
        zF_s(\bq)-1 = - D_0 \bq
            \cdot\Biggl[ \frac{1 + (1+\gamma^2)K_{0,22}}{(1+K_{0,11})(1+K_{0,22}) + \gamma^2 K_{0,11}K_{0,22}}\be_1\otimes\be_1 + \frac{1}{1+K_{0,33}}\be_3\otimes\be_3 \Biggr]\cdot\bq F_s(\bq, z).
    \end{equation}
This equation of motion is formally similar to Eq.~\eqref{eq:zphi-1}, with the collective memory kernel being replaced by the tracer's memory kernel. For a wavevector $\bq$ lying in the $(xy)$ plane, we get, using an inverse Laplace transform
\begin{equation}
    \p_t F_s + D_0q^2 F_s + D_0 q^2(1+\gamma^2)K_{0,22}* F_s 
    = -\left[\left(K_{0,11}  + K_{0,22}\right) +  (1+\gamma^2)K_{0,11}*K_{0,22}\right]*\p_tF_s,
\end{equation}
where the dependence on $(\bq,t)$ has been omitted for clarity. This equation was announced in an earlier work (Eq.~(11) in \cite{ghimenti2023sampling}). The properties of this dynamics can be analyzed in the same way illustrated for the dynamical structure factor in Sec.~\ref{sec:schematic_new}. 

\subsection{Diffusion tensor}

The diffusion tensor at any finite wavevector and frequency is obtained by substituting the mode-coupling expression for $\bK_0$, given in Eq.~\eqref{eq:K0_mct}, into Eq.~\eqref{eq:diffusivity_tensor_projection}. The result reads, in the $\bA-\bq$ basis introduced in Eq.~\eqref{eq:Aqbasis}
    \begin{equation}\label{eq:Dtensor_mct}
        \begin{split}
            & \bD(\bq,z) = D_0\Biggl[ \frac{1 + (1+\gamma^2)K_{0,22}}{(1+K_{0,11})(1+K_{0,22}) + \gamma^2 K_{0,11}K_{0,22}}\be_1\otimes\be_1 + \frac{1}{1+K_{0,33}}\be_3\otimes\be_3 \\
            &+ \gamma \frac{K_{0,11} + K_{0,22} + (1+\gamma^2)K_{0,11}K_{0,22}}{(1+K_{0,11})(1+K_{0,22}) + \gamma^2 K_{0,11}K_{0,22}}\left[\be_1\otimes\be_2-\be_2\otimes\be_1 \right] 
            + \frac{(1+K_{0,11})K_{0,22} + K_{0,11}(K_{0,22}-1)\gamma^2}{(1+K_{0,11})(1+K_{0,22}) + \gamma^2 K_{0,11}K_{0,22}} \be_2\otimes\be_2\Biggr].
         \end{split}
    \end{equation}
\end{widetext}
The expression of the diffusion tensor is one of the main results of this work. From the large wavelength, small frequency limit of Eq.~\eqref{eq:Dtensor_mct} we obtain the diffusion matrix $\bD$. In Cartesian coordinates, it reads
\begin{equation}
    \begin{split}
       \bD &= D_{\parallel,x}\left[\be_x\otimes\be_x + \be_y\otimes\be_y\right] \\
       &+ D_{\parallel,z}\be_z\otimes\be_z \\
       &+ D_\perp\left[\be_y\otimes\be_x - \be_z\otimes\be_y\right].
    \end{split}
\end{equation}
We now discuss the expression and the behavior of the different diffusion constants that appear in the diffusion tensor $\bD$.

\subsubsection{Longitudinal diffusion}

The dynamics with transverse forces is anisotropic. As a consequence, there are two types of longitudinal diffusion constants, one related to the diffusion in the $(xy)$ plane (within which rotational invariance is preserved), $D_{\parallel,x}$ and one related to the diffusion along the $z$ direction, $D_\parallel,z$. Their expressions are
\begin{equation}
    \begin{split}
        D_{\parallel,x}(\gamma) &= D_0\frac{1 + (1+\gamma^2)K_{0,11}^{\infty}}{(1+K_{0,11}^{\infty})(1+K_{0,22}^{\infty}) + \gamma^2 K_{0,11}^{\infty}K_{0,22}^{\infty}} \\
        D_{\parallel,z}(\gamma) &= D_0\frac{1}{1+K_{0,33}^{\infty}} ,
    \end{split}
\end{equation}
where we introduced the notation $K_{0,ii}^\infty\equiv K_{0,ii}(\bq\to0,z\to0)$ to denote the long wavelength, zero frequency of the tracer's memory kernel. This coefficient reads explicitly
\begin{equation}
    K_{0,ii}^{\infty} = \frac{D_0}{\rho_0} \int_0^{+\infty}\dd t \int_\bk  [k_i\rho_0 c(k)]^2 F_s(-\bk,t)S(\bk,t).
\end{equation}

For $\gamma=0$, the equilibrium result is duly recovered:
\begin{equation}
    D_{aa,\text{\tiny{eq}}} = D_0\frac{1}{1+K_{0,\parallel,\text{\tiny{eq}}}^{\infty}} ,
\end{equation}
for $a=x,y,z$. When $\gamma\neq0$ the anisotropy of the dynamics is revealed through the different expressions of the diffusion constants. 

The dynamics is indeed accelerated. If we assume that in the ergodic phase $K_{0,ii}^{\infty}\leq K_{0,ii,\text{\tiny{eq}}}^\infty$, we self-consistently obtain $D_{aa}(\gamma)\geq D_{aa}(0)$. 

In the ergodic phase in the limit of strong drive, the memory kernel scales as $\gamma^{-1}$. This implies that the efficiency of transverse forces with respect to the equilibrium dynamics, computed as the ratio $D(\gamma)\equiv\sum_{a=x,y,z}D_{aa}(\gamma)$ over $D(0)$, grows as
\begin{equation}
    \frac{D(\gamma)}{D(0)} \sim \gamma.
\end{equation}

We now investigate two limiting cases: in the high temperature regime, the memory kernels vanish because the fluid enters an effectively noninteracting limit and therefore $D(\gamma)=D_0$, while at $T=T_{\text{MCT}}$ the memory kernels diverge as predicted by the schematic approach of Sec.~\ref{schematic}, implying that $D(\gamma)=0$ and dynamical arrest occurs. We have therefore
\begin{equation}
  \begin{aligned}
    \lim_{T\to\infty}\frac{D(\gamma)}{D(0)} &= c(\gamma), \\
    \lim_{T\to T_{\text{MCT}}}\frac{D(\gamma)}{D(0)} &= 1, 
  \end{aligned}
\end{equation}
With $c(\gamma)>1$ a constant that depends on the intensity of the driving force. These conditions, together with the acceleration in the ergodic phase, demonstrate that the ratio of the diffusion constants $\frac{D(\gamma)}{D_0}$ exhibits a maximum as a function of the temperature in the ergodic phase. This maximum is observed in numerical simulations~\cite{ghimenti2023sampling}. 

\subsubsection{Odd diffusivity}

The odd diffusion constant is encoded in the antisymmetric part of the diffusivity tensor:
\begin{equation}
    D_\perp(T,\gamma) =  -D_0\gamma \frac{K^{\infty}_{0,11} + K^{\infty}_{0,22} + (1+\gamma^2)K^{\infty}_{0,11}K^{\infty}_{0,22}}{(1+K^{\infty}_{0,11})(1+K^{\infty}_{0,22}) + \gamma^2 K^{\infty}_{0,11}K^{\infty}_{0,22}} .
\end{equation} 
For an asymptotically large driving, $\gamma\to \infty$, the odd diffusion constant grows linearly in $\gamma$, $D_\perp(\gamma,T)\sim \gamma$. This suggests that strong transverse forces manifest themselves with an increasingly swirling motion.

Close to ergodicity breaking where the memory kernel diverges, we obtain
\begin{equation}
  \lim_{T\to T_{\text{MCT}}} D_\perp = -\gamma D_0.
\end{equation}
Therefore, even if the longitudinal diffusion constant associated to particle transport goes to $0$ at the critical temperature, the odd diffusion constant remains nonzero. Together with the vanishing of the diffusion constant, this suggests a physical picture where particles perform a swirling motion inside the permanent local cage made by their neighbors. An analogous situation can be shown to arise in the much simpler model of a particle in an harmonic well under the action of transverse forces~\cite{ghimenti2023sampling}.   

\subsection{Mobility tensor}

The mobility tensor is an alternative quantity characterizing the nature of the dynamics with a physical content distinct from that of the diffusivity tensor due to the breaking of the Einstein relation. The mobility tensor is obtained from Eq.~\eqref{eq:mobility_tensor_projection}, using the mode-coupling expansion of the tracer's memory kernel given in Eq.~\eqref{eq:K0_mct}. The result reads
\begin{widetext}
\begin{equation}
    \begin{split}
        \bmu(\bq,z) &= D_0\beta\Biggl[ \frac{1+K_{22}}{(1+K_{11})(1+K_{22}) + \gamma^2 K_{11} K_{22}}\be_1\otimes\be_1 + \frac{1}{1+K_{33}}\be_3\otimes\be_3 \\
        &+ \frac{\gamma\left(K_{0,11}\be_2\otimes\be_1 - K_{0,22}\be_1\otimes\be_2\right)}{(1+K_{11})(1+K_{22}) + \gamma^2 K_{11} K_{22}} 
        + \frac{1+K_{11}}{(1+K_{11})(1+K_{22}) + \gamma^2 K_{11} K_{22}}\be_2\otimes\be_2\Biggr] .
    \end{split}
\end{equation}
\end{widetext}
In the small frequency and large wavelength limit of these expression, we obtain the mobility of the tracer,
\begin{equation}
    \begin{split}
        \bmu &= \mu_{\parallel,x}\left[\be_x\otimes\be_x + \be_y\otimes\be_y\right] + \mu_{\parallel,z}\be_z\otimes\be_z \\
        &+ \mu_\perp \left[\be_x\otimes\be_y - \be_y\otimes\be_x\right] ,
    \end{split}
\end{equation}
with the longitudinal mobilities $\mu_{\parallel,x}$,$\mu_{\parallel,z}$, and odd mobility $\mu_\perp$ given by
\begin{equation}
    \begin{split}
        \mu_{\parallel,x} &= D_0\beta \frac{1+K_{0,11}^{\infty}}{(1+K_{0,11}^\infty)^2 + \gamma^2 K_{0,11}^\infty} ,\\
        \mu_{\parallel,z} &= D_0\beta \frac{1}{1+K_{0,33}^\infty} ,\\
        \mu_{\perp} &= -\gamma D_0\beta \frac{K_{0,11}^{\infty}} {(1+K_{0,11}^\infty)^2 + \gamma^2 K_{0,11}^\infty} .
    \end{split}
\end{equation}
The anisotropic character of transverse forces in three dimensions manifests itself through the fact that $\mu_{\parallel,z}\neq\mu_{\parallel,x}$. In the large $\gamma$-limit, all nonzero entries of the mobility tensor converge (in modulus) to the mobility of a free tracer, $D_0\beta$. This is in contrast with the behavior of the diffusivity tensor, whose nonzero entries grow linearly in $\gamma$. This is a consequence of the breakdown of the fluctuation-dissipation theorem, arising from the nonequilibrium nature of the dynamics. A variant of the fluctuation-dissipation theorem survives for the longitudinal components of the diffusion and mobility tensor, as $[\mu (\mathbf{1}+\gamma\bA)]_{aa}=\beta[\bD]_{aa}$ with $a=x,y,z$. This variant of the fluctuation-dissipation theorem was also derived in the dynamical mean-field treatment of transverse forces~\cite{ghimenti2024transverse}. 

As the glass transition is approached, all components of the mobility tensor vanish. Physically, this means that in the dynamically arrested glass, a weak external force cannot set the particles in motion, neither in the longitudinal nor in the transverse direction with respect to the external force. 

The vanishing of longitudinal and odd mobilities is one of the facets of the glass transition in the mode-coupling theory of transverse forces. Another facet is the divergence of the viscosity tensor, which will be explored  in the next section.

\section{Odd viscosity}

This section is devoted to the study of the emerging odd viscosity~\cite{avron1998odd, banerjee2017odd, epstein2020time, hargus2020time,  fruchart2023odd}. Following \cite{vogel2020stress, epstein2020time}, we compute the viscosity tensor in the long wavelength and long times limits by looking at the response of the stress to an external perturbation. To do so, we define a microscopic stress tensor in the presence of transverse forces and develop its linear response theory, leading to the definition of the viscosity tensor, and in particular to the identification of its odd component.

\subsection{Stress tensor and transverse forces}

Following \cite{vogel2020stress}, we start by defining the microscopic stress tensor in the presence of transverse forces from the equation
\begin{equation}
\label{eq:def_stress}
  \partial_t n(\bq) =  \Omega^\dagger_\gamma n(\bq) \equiv -D_0\beta q_aq_b \sigma_{ab}(\bq).
\end{equation}
This identity, which only holds when inserted into the average $\langle\ldots\rangle$, yields
\begin{equation}\label{eq:sigma}
   \sigma_{ab}(\bq) \equiv \sum_i \left[i(\delta_{bc} + \gamma A_{bc})\frac{q_aF_{i,c}}{q^2} + T\delta_{ab}\right]\ee^{-i\bq\cdot\br_i}.
\end{equation}
For $\gamma=0$, we recover the expression of the microscopic equilibrium stress tensor, $\sigma^{\text{eq}}(\bq)$, given by
\begin{equation}
  \sigma^{\text{eq}}_{ab}(\bq) \equiv \sum_i \left[i\frac{q_aF_{i,b}}{q^2} + T\delta_{ab}\right]\ee^{-i\bq\cdot\br_i}.
\end{equation}
For equilibrium dynamics, the viscosity is defined as the response function of $\mQ\sigma^{\text{eq}}$ to particle current fluctuations, as discussed in \cite{hess1981fokker} for underdamped Langevin dynamics and extended in \cite{vogel2020stress} for the overdamped case. In the presence of transverse forces one must substitute $\sigma^{\text{eq}}$ with $\sigma$ given by Eq. \eqref{eq:sigma}. This gives rise to coupling between parallel and longitudinal components of the stress tensor, which eventually results in odd viscosity. 

Note that the projection of the stress tensor in the space orthogonal to the density fluctuations, $\mQ\sigma_{ab}(\bq)$, is related with its equilibrium counterpart by the following relation
\begin{equation}
    \mQ\sigma_{ab}(\bq) = \mQ\sigma_{ac}^{\text{eq}}(\delta_{cb} - \gamma A_{cb}),
\end{equation}
where the Einstein summation convention is being used. We also have 
\begin{equation}\label{eq:j_as_sigma}
    \mQ j_a(\bq) =-iq_b\mQ \sigma_{ba}^{\text{eq}}(\bq) ,
\end{equation}
with $\mQ j_a(\bq)$ the projected force density Fourier modes defined in Eq. \eqref{eq:Qj}.

In the next section, we pursue our program by developing a linear response theory of the microscopic stress.

\subsection{Linear response theory}

We consider the perturbation produced by a weak external velocity field made by a single Fourier mode $\bv(\br,t) \equiv \frac{1}{V}\bv(t)\ee^{i\bq\cdot\br}$. The evolution operator associated to the system is now $\Omega_\gamma + \delta\Omega$, with
\begin{equation}
  \begin{aligned}
    \delta\Omega &\equiv - \frac{1}{V}\sum_i \bnabla_i \cdot \bv(t)\ee^{i\bq\cdot\br_i} .
  \end{aligned}
\end{equation}
As a consequence of the perturbation, the probability distribution associated to the system becomes $\rho_\text{\tiny{B}}(\br^N) + \delta\rho(\br^N,t)$. The equation of motion for the perturbation $\delta\rho(\br^N,t)$ reads, to linear order,
\begin{equation}
  \partial_t\delta\rho(\br^N,t) = \Omega_\gamma\delta\rho(\br^N,t) + \delta\Omega(t)\rho_\text{\tiny{B}}(\br^N).
\end{equation}
The solution to this equation is given by
\begin{equation}
  \begin{aligned}
    &\delta\rho(\br^N,t) = \int_{-\infty}^t d\tau \ee^{\Omega_\gamma(t-\tau)}\delta\Omega(\tau)\rho_\text{\tiny{B}}(\br^N) \\   
    &= -\frac{\beta}{V} \int_{-\infty}^t \dd\tau \ee^{\Omega_\gamma(t-\tau)}v_b(\tau)iq_a\sigma^{\text{eq}*}_{ab}(\bq)\rho_{\text{\tiny{B}}}(\br^N) .
    \end{aligned}
\end{equation}
The average of the stress tensor in linear response reads
\begin{equation}\label{sigmalr}
  \begin{aligned}
  \left\langle \mQ \sigma_{ab}(\bq,t)\right\rangle^{\text{lr}} &= -\frac{\beta}{V}\int_{-\infty}^t
  \dd\tau iq_cv_d(\tau) \\
  &\times \left\langle \sigma^{\text{eq}*}_{cd}(\bq)\ee^{\Omega_{-\gamma}(t-\tau)}\mQ \sigma_{\alpha\beta}(\bq)\right\rangle \\
  &= -\frac{\beta}{V}\int_{-\infty}^{t}
  \dd\tau iq_c v_d(\tau) \\ 
  &\times \left\langle \sigma^{\text{eq}*}_{cd}(\bq)\ee^{\Omega_{-\gamma}(t-\tau)}\mQ\sigma_{ab}(\bq)\right\rangle\theta(t-\tau) .
  \end{aligned}
\end{equation}
The average $\langle \ldots \rangle^{\text{lr}}$ is an average over the dynamics described by the operator $\Omega_\gamma + \delta\Omega$, neglecting terms of order higher than linear in $\bv(t)$.  We are interested in computing the variation of the stress with respect to the small perturbation. To do this, it is more suitable to work in Fourier space, by denoting the Fourier transform of a function $f(t)$ as $f(\omega)\equiv \int_{-\infty}^{+\infty} f(t)\ee^{-i\omega t}$. Within the linear response framework, we can take the perturbation $\bv$ to be a monochromatic plane wave, i.e. $\bv(t) = \bv(\omega)\ee^{-i\omega t+0_+t}$. The symbol $0_+$ has to be understood as a small positive infinitesimal quantity used to ensure that the perturbation goes to $0$ at $t=-\infty$. Substitution into Eq.~\eqref{sigmalr} yields
\begin{equation}
  \frac{\partial\left\langle \mQ\sigma_{ab}(\bq,\omega)\right\rangle^{\text{lr}}}{\partial iq_c v_d} = \frac{\beta}{V} \left\langle \sigma^{\text{eq}*}_{cd}(\bq)\frac{1}{-i\omega - \Omega_{-\gamma}}\mQ\sigma_{ab}(\bq) \right\rangle.
\end{equation}
This is the response of $\mQ\sigma_{ab}(\bq)$ to an external velocity gradient. To obtain the viscosity, we need to work out the response of $\mQ\sigma_{ab}(\bq)$ to a change in the gradients of the particle current. However, within linear response, we can work out a relation between the particle currents and the applied velocity field, thus making the calculation of the viscosity tensor possible. This is done next.

\subsection{The viscosity tensor}

Following ~\cite{hess1981fokker,  cichocki1987memory, vogel2020stress}, we introduce the average particle current  $\bJ(\bq,t)$, governing the evolution of the density mode $n(\bq)$
\begin{equation}\label{eq:def_pc}
  \left\langle \partial_t n(\bq,t) \right\rangle^{\text{lr}} \equiv i\rho_0\bq\cdot\bJ(\bq,t).
\end{equation}
The left hand side of Eq. \eqref{eq:def_pc} splits into
\begin{equation} 
  \left\langle \partial_t n(\bq,t) \right\rangle^{\text{lr}} = \left\langle n(\bq,t)\delta\Omega(t)\right\rangle + \left\langle n(\bq,t)\Omega_{\gamma}\right\rangle^{\text{lr}} ,
\end{equation}
which implies
\begin{equation}\label{j_a}
  J_a(\bq,t) = v_a(\bq,t) - \frac{D_0\beta}{\rho_0}q_b\left\langle \sigma_{ab}(\bq,t)\right\rangle^{\text{lr}}.
\end{equation}
Thus, the particle current corresponds to sum of the imposed solvent velocity $\bv$ and of a term that originates from the change in the stress tensor.

The viscosity tensor is defined as the response function describing the change of the projected stress tensor due to the gradient of the average particle current~\cite{hess1981fokker}:
\begin{equation}\label{eq:eta}
  \eta_{abcd}(\bq,\omega) \equiv \frac{\partial\left\langle \mQ\sigma_{ab}(\bq,\omega)\right\rangle^{\text{lr}}}{\partial iq_c J_d(\bq,\omega)}.
\end{equation}

Our aim is now to show that the viscosity tensor is related to the following correlation function,
\begin{equation}
    C^{\text{irr}}_{abcd}(\bq,\omega) \equiv \frac{\beta}{V}\left\langle \sigma^{\text{eq}*}_{ab}(\bq) \mQ\frac{1}{-i\omega - \Omega^{\text{irr}}_{-\gamma}}\mQ \sigma_{cd}(\bq) \right\rangle,
\end{equation}
which encodes the correlation between the equilibrium stress tensor and the stress tensor in presence of transverse forces, evolving with the irreducible operator $\Omega^\text{irr}_{-\gamma}$ defined in Eq.~\eqref{eq:Omegairr}. 

We first introduce an auxiliary correlation function $C^\mQ_{abcd}$,
\begin{equation}
    C^{\mQ}_{abcd} \equiv \frac{\beta}{V} \left \langle\sigma_{ab}^{\text{eq}*}(\bq) \mQ\frac{1}{-i\omega - \mQ\Omega_{-\gamma}\mQ}\mQ\sigma_{cd}(\bq) \right\rangle ,
\end{equation}
which is the analogue of $C^{\text{irr}}_{abcd}$ but evolves with the projected evolution operator $\mQ\Omega_{-\gamma}\mQ$. Using the operator identity \begin{equation}
    \begin{split}
        \frac{1}{-i\omega - \mQ\Omega_{-\gamma}} &= \frac{1}{-i\omega - \Omega_{-\gamma}} \\
        &+ \frac{1}{-i\omega - \Omega_{-\gamma}}\mP\Omega_{-\gamma}\frac{1}{-i\omega - \mQ\Omega_{-\gamma}} ,
    \end{split}
\end{equation}
we obtain
\begin{equation}\label{eq:Qsigma_as_CQ}
    \begin{split}
       \frac{\partial \langle \mQ\sigma_{ab}(\bq)\rangle^\text{lr}}{\partial iq_c v_d} &= \Biggl[ \delta_{ce}\delta_{dg} \\
       &- \frac{D_0\beta}{S(q)}q_f\frac{\p\langle n(\bq,t)\rangle^{\text{lr}}}{\partial iq_cv_d}(\delta_{fg} + \gamma A_{fg})q_e\Biggr] C^\mQ_{egab}.
    \end{split}
\end{equation}
On the other hand, the correlators $C^\mQ_{abcd}$ and $C^\text{irr}_{abcd}$ are related to each other. We can exploit the resolvent identity given by Eq.~\eqref{eq:resolventid_irr} and the fact that the operator $\delta\Omega_{-\gamma}$ defined in Eq.~\eqref{eq:deltaOmega} can be expressed in terms of the stress tensor through Eq. \eqref{eq:j_as_sigma},
\begin{equation}
    \delta\Omega_{-\gamma} = \frac{D_0\beta^2}{N}q_a q_d \mQ\sigma^{\text{eq}}_{ab}(\bq)\rangle[\delta_{bc} - \gamma A_{bc}]\langle \sigma_{dc}^{\text{eq}}\mQ.   
\end{equation}
This yields
\begin{equation}\label{eq:Cirr_as_C}
    \begin{split}
        C^\text{irr}_{abcd} &= C^{\mQ}_{abcd} - \frac{D_0\beta^2}{\rho_0V}q_e q_f \\
        &\times\langle\sigma^{\text{eq}*}_{ab}(\bq) \frac{1}{-i\omega-\mQ\Omega_{-\gamma}\mQ}\mQ\sigma^{\text{eq}}_{ef}(\bq)\rangle(\delta_{fg} - \gamma A_{fg})C^{\text{irr}}_{fgcd} \\
        &= C^{\mQ}_{abcd} - \frac{D_0\beta}{\rho_0}q_e q_f C^\mQ_{abeg}C^{\text{irr}}_{fgcd}. 
    \end{split}
\end{equation}
This equation is analogous to the relation between the memory kernels $\bK$ and $\bKtilde$ given by Eq.~\eqref{eq:Ktilde_K}. Contracting both sides of Eq.~\eqref{eq:Cirr_as_C} with the quantity inside the brackets in Eq.~\eqref{eq:Qsigma_as_CQ} and rearranging the terms gives
\begin{equation}\label{eq:j_Cirr_first}
    \begin{split}
         \frac{\partial \langle \mQ\sigma_{ab}(\bq)\rangle^\text{lr}}{\partial iq_c v_d} &= \Biggl[ \delta_{ce}\delta_{df} + \frac{D_0\beta}{\rho_0}q_e q_g \frac{\partial \langle \mQ\sigma_{gf}(\bq)\rangle^{\text{lr}}}{\partial iq_c v_d}  \\
         &- q_eq_g\frac{D_0}{\rho_0S(q)}\frac{\partial \langle n(\bq)\rangle^{\text{lr}}}{\partial iq_c v_d}(\delta_{gf} + \gamma A_{gf})\Biggr] C^{\text{irr}}_{efab}.
    \end{split}
\end{equation}
On the other hand, from Eq.~\eqref{j_a} we have 
\begin{equation}\label{eq:dj_dv}
    \begin{split}
        \frac{\partial iq_e J_f}{\partial iq_c v_d} &= \delta_{ce}\delta_{df} + \frac{D_0\beta}{\rho_0}q_e q_g \frac{\partial \langle \mQ\sigma_{hf}(\bq)\rangle^{\text{lr}}}{\partial iq_c v_d}  \\
        &- q_eq_g\frac{D_0}{\rho_0S(q)}\frac{\partial \langle n(\bq)\rangle^{\text{lr}}}{\partial iq_c v_d}\delta_{gf} .
    \end{split}
\end{equation}
Combining Eqs.~(\ref{eq:j_Cirr_first}, \ref{eq:dj_dv}), using the chain rule and the definition of the viscosity tensor given by Eq.~\eqref{eq:eta} we obtain
\begin{equation}\label{eq:eta_Cirr}
    \begin{split}
        \frac{\partial iq_e J_f}{\p iq_c v_d}\eta_{abef}(\bq,\omega) &= \Biggl[  \frac{\partial iq_e J_f}{\p iq_c v_d} \\
        &- \gamma q_eq_g\frac{D_0}{\rho_0S(q)}\frac{\partial \langle n(\bq)\rangle^{\text{lr}}}{\partial iq_c v_d} A_{gf}\Biggr] C^{\text{irr}}_{efab}(\bq,\omega)   .
    \end{split}
\end{equation}
This equation relates the irreducible stress-stress correlator with the viscosity. For $\gamma=0$, we recover the relation derived in~\cite{vogel2020stress},
\begin{equation}
    \eta_{abcd}(\bq,\omega) =\frac{\beta}{V}\langle \sigma^{\text{eq}*}_{cd}(\bq)\frac{1}{-i\omega - \Omega^{\text{irr}}_0}\mQ\sigma^{\text{eq}}_{ab}(\bq)\rangle.
\end{equation}
When $\gamma\neq 0$, and additional term stemming from the influence of the transverse forces and proportional to the response of the density field to external currents appears. Moreover, the stress-stress correlator $C^{\text{irr}}$ depends on $\gamma$ both through the definition of the stress $\sigma_{ab}$ and the dynamical evolution operator $\Omega_{-\gamma}^{\text{irr}}$.

In  the hydrodynamic limit, when $q\to 0$ and $\omega\to 0$, the second term on the right hand side of Eq.~\eqref{eq:eta_Cirr} is subleading compared to the first one, leading to
\begin{equation}\label{eq:viscosity_formal}
    \eta_{abcd}(\bq\to0,\omega\to0) = C^{\text{irr}}_{cdab}(\bq\to0, \omega\to0).
\end{equation}
This equation relates the hydrodynamic viscosity with the stress-stress irreducible correlator. In the next two sections, we use this formula to compute, within the mode-coupling approximation, the shear and odd viscosities. 

\subsection{Shear viscosity}

We start with the shear viscosity $\eta_{xyxy}$. Due to the rotational invariance of the dynamics in the $(xy)$ plane, we evaluate the $\bq\to0$ limit with $\bq=q\be_x$. From Eq.~\eqref{eq:viscosity_formal} we obtain
\begin{equation}
    \begin{split}
        \eta_{xyxy} &= \lim_{q\to0}\lim_{\omega\to0} \frac{\beta}{V}\left\langle\sigma^{\text{eq}*}_{xy}(\bq)\mQ\frac{1}{-i\omega - \Omega_{-\gamma}^{\text{irr}}}\mQ\sigma_{xy}^{\text{eq}}(\bq)\right\rangle \\
        &-\gamma\frac{\beta}{V}\left\langle\sigma^{\text{eq}*}_{xy}(\bq)\mQ\frac{1}{-i\omega - \Omega_{-\gamma}^{\text{irr}}}\mQ\sigma_{xx}^{\text{eq}}(\bq)\right\rangle.
    \end{split}
\end{equation}
We now replace $\mQ \sigma_{ab}^{\text{eq}}(\bq)$ with $i\mQ \frac{q_a}{q^2}j_b(\bq)$. In this way, the stress-stress correlations is expressed in terms of the projected force-force correlations, encoded by the memory kernel $\bK$ given by Eq. \eqref{eq:K}. We therefore get, using the $\bA-\bq$ basis,
\begin{equation}\label{eq:shearviscosity}
        \eta_{xyxy} = \lim_{q\to0}\lim_{\omega\to0} \frac{\rho_0}{D_0\beta} \left[K_{22}(q\be_x,-i\omega) -\gamma K_{21}(q\be_x,-i\omega) \right]. 
\end{equation}
Using the expression of $\bK$ given by the mode-coupling approximation, Eq.\eqref{eq:Kij}, and the fact that within this approximation the off-diagonal terms in $\bK$ are $0$ in the $\bA-\bq$ basis, we have
\begin{equation}
    \begin{split}
        \eta_{xyxy} &= \lim_{q\to0}\lim_{\omega\to0} \frac{\rho_0}{q^2 D_0\beta} K_{22}(q\be_x,-i\omega) \\
        &=  \lim_{q\to 0}\frac{\rho_0^2}{2q^2\beta}\int_0^{+\infty}\dd t \int_\bk [k_y (c(k) - c(|q\be_x-\bk|)) ]^2 \\
        &\times S(\bk,t)S(q\be_x-\bk,t) \\
        &= \frac{1}{2\beta}\int_0^{+\infty} \dd t \int_\bk [\khat_xk_y\rho_0c'(k)]^2 \lvert S(\bk,t)\rvert^2
    \end{split} .
\end{equation}
In this expression, the dependence on $\gamma$ enters through the dynamical evolution of $S(\bk,t)$. For $\gamma=0$, rotational symmetry is restored, and we recover the equilibrium, mode-coupling result \cite{nagele1998linear}:
\begin{equation}
    \eta_{xyxy}^{\text{eq}} = \frac{1}{60\pi^2\beta}\int_0^{+\infty} \dd k k^4 \left[ \frac{1}{S(k)}S'(k)\right]^2\phi(k,t)^2.
\end{equation} 
Since the relaxation of $S(\bk,t)$ is faster in the presence of transverse forces we have $\eta_{xyxy}(\gamma)\leq\eta_{xyxy}^{\text{eq}}$. Also, the shear viscosity diverges at the glass transition. 

\subsection{Odd viscosity}

\newcommand{\Sigmairr}{\Sigma^{\text{irr}}}

We now address the odd viscosity. We identify the following contribution as the odd viscosity:
\begin{equation}\label{etao}
  \eta_{\text{odd}} \equiv \frac{1}{2}\left(\eta_{xyxx} - \eta_{xxxy}\right).
\end{equation}
Physically, a nonzero $\eta_\text{odd}$ means that attempts at compressing the system along the $x$-direction generate shear flows in the $(xy)$ plane. In a nonreciprocal fashion, shear stresses applied on the $(xy)$ plane will generate an expansion of the liquid along the $x$-direction. In equilibrium dynamics we have $\eta_{abcd}=\eta_{cdba}$ and therefore $\eta_\text{odd}=0$. We address now how this situation changes when transverse forces are present.

Performing a computation similar to the one done in Eq.~\eqref{eq:shearviscosity} gives
\begin{widetext}
\begin{equation}\label{eq:etaodd_before_mct}
    \eta_{\text{odd}} = \frac{1}{2} \lim_{q\to0}\lim_{\omega\to0} \frac{\rho_0}{q^2 D_0\beta} \left[ (K_{21}(q\be_x,-i\omega) + K_{12}(q\be_x,-i\omega)) - \gamma (K_{11}(q\be_x,-i\omega) + K_{22}(q\be_x,-i\omega))\right].
\end{equation}
Within the mode-coupling approximation, the first two terms on the right hand side of Eq. \eqref{eq:etaodd_before_mct} are $0$. Taking the hydrodynamic limit of the mode-coupling expression of $K_{ii}$ given by Eq.~\eqref{eq:Kij}, we obtain
\begin{equation}
    \begin{split}
        \eta_{\text{odd}} &= -\frac{\gamma}{2} \lim_{q\to0}\lim_{\omega\to0} \frac{\rho_0}{q^2 D_0\beta} \left[ K_{11}(q\be_x,-i\omega) + K_{22}(q\be_x,-i\omega)\right] \\
        &= -\frac{\gamma}{4\beta}\int_0^{+\infty}\dd t\int_\bk \left[[\rho_0c(k) + k_x\khat_x \rho_0 c'(k)]^2 + [\hat{k}_x k_y \rho_0c'(k)]^2\right]S(\bk,t)^2 .
    \end{split}
\end{equation}
\end{widetext}
As the glass transition is approached the odd viscosity diverges. Physically, this means that in the glassy state no form of transport is possible, and the transverse, as well as the longitudinal, dynamical pathways through which the liquid relaxes, are blocked. This directly impacts all viscosity coefficients. 

\section{Lifting: sampling the Boltzmann distribution with active particles}

\label{sec:MCTlifting}

The term lifting is used in the Markov Chain Monte Carlo literature to denote a sampling technique used to extract configurations from a given steady state target distribution, see \cite{vucelja2016lifting} for a pedagogical review. In a lifting scheme, the number of degrees of freedom of the system is extended, and transitions involving the additional degrees of freedom are allowed. By exploiting the extended nature of the phase space, one can perform irreversible transitions, while preserving the target distribution in the steady state. The nonequilibrium nature of the dynamics can be exploited to obtain a convergence towards the equilibrium steady state which is faster than conventional equilibrium methods, such as the Metropolis-Hastings algorithm.

Lifted schemes have been employed in a variety of contexts: in the Ising model in mean-field~\cite{turitsyn2011irreversible, monthus2021large} and in one-dimension~\cite{sakai2013dynamics}, in hard spheres in one~\cite{kapfer2017irreversible} and two dimensions (where it is named Event Chain Monte Carlo method~\cite{bernard2009event, michel2014generalized}), in lattice random walkers~\cite{sakai2016eigenvalue}, and glassy hard disks~\cite{ghimenti2024irreversible}. Based on these examples, it appears that the performance is remarkable even when the system evolves in a not-too-strongly convex potential. This is what occurs when the lifted degrees of freedom are coupled with the order parameter close to a second-order phase transition.

Lifted dynamics can be described in terms run and tumble particles with a properly tuned tumbling rate. By exploiting this connection, we now present a model of lifted dynamics inspired by Active Brownian Particles (ABPs). While the mode-coupling theory of the non-lifted counterpart of these active system has been explored recently~\cite{szamel2016theory, liluashvili2017mode, szamel2019mode}, here we focus on systems of active particles specifically designed to sample the Boltzmann distribution, and we compare this model with the overdamped equilibrium dynamics counterpart. The results obtained will also make clear that the performance of lifted schemes in supercooled liquids is similar to the one obtained through the transverse forces approach studied in the first part of this article, thus strengthening the connection between lifting and transverse forces.

\subsection{Lifted active Brownian particles (ABP)}

The lifted-ABP dynamics of $N$ particles in two dimensions is given by
\begin{equation}
\label{eq:lABPs}
  \begin{cases}
    \dot\br_i &= v_0 \buhat_i  \\
    \dot\theta_i &= v_0\beta \bF_i\cdot \bA\buhat_i + \sqrt{2D_r}\chi_i ,
  \end{cases}
\end{equation}
where $\buhat_i=(\cos{\theta_i}, \sin{\theta_i})$ is a unit vector denoting the direction of the self-propulsion speed of particle $i$. The vector $\buhat_i$ forms an angle $\theta_i$ with respect to the $x$-axis. $\chi_i$ is a Gaussian white noise, $\langle \chi_i(t) \chi_j(t')\rangle = \delta_{ij}\delta(t-t')$. The self-propulsion orientation is subjected to diffusion with a diffusion constant $D_r$ as for standard ABPs, but is in addition subjected to a drift that depends on the interaction force $\bF_i = -\sum_i \bnabla_{j\neq i} V(\br_i-\br_j)$ between the particles in position space. The latter enforces the stationarity of the Boltzmann distribution in the steady state because the matrix $\bA \equiv \begin{bmatrix} 0 & -1 \\ 1 & 0 \end{bmatrix}$ ensures that the drift is always perpendicular to $\buhat_i$. The drift term makes the ABP equations of motion non-conventional, hence the `lifted-ABP' name for Eq.~\eqref{eq:lABPs}.  

In the absence of interactions when $\bF_i=0$, Eq.~\eqref{eq:lABPs} describes freely diffusing ABPs, with diffusion constant $\frac{v_0^2}{2D_r}$. When interactions are added, lifted-ABPs sample the equilibrium Boltzmann distribution of the interacting system at fixed temperature $T$, and this can be done at various values of $\frac{v_0^2}{2D_r}$. This is to be compared with the equilibrium dynamics where temperature simultaneously fixes the Boltzmann distribution and the free diffusion constant $D_0 = \mu_0 T$. 
  
The evolution of the probability distribution $\rho(\br^N,\theta^N,t)$ of lifted-ABPs is governed by the operator $\Omega_{v_0}$:
\begin{equation}
    \begin{split}
      \Omega_{v_0} \equiv & - \sum_i \bnabla_i \cdot v_0\buhat_i \\
      &+\sum_i \partial_{\theta_i} \left[  -v_0\beta \bF_i\cdot \bA \buhat_i + D_r\partial_{\theta_i}\right] ,
    \end{split}
\end{equation}
so that
\begin{equation}\label{ABPpartialrho=Omegav0}
  \partial_t\rho(\br^N,\theta^N,t) = \Omega_{v_0}\rho(\br^N,\theta^N,t).
\end{equation}

The steady state solution of Eq.~\eqref{ABPpartialrho=Omegav0} is a product of independent distributions in the respective subspaces spanned by $\br^N$ and by $\bu^N$:
\begin{equation}
  \rho_{\text{ss}} = \frac{1}{(2\pi)^N}\rho_{\text{\tiny{B}}}(\br^N) = \frac{1}{Z}e^{-\beta \sum_{i<j}V(\br_i-\br_j)}.
\end{equation}
with $Z\equiv (2\pi)^N\int d\br^Ne^{-\beta \sum_{i<j}V(\br_i-\br_j)}$.

The operator $\Omega_{v_0}$ has the following property, which is a reflection of the breaking of detailed balance reminiscent of Eq.~\eqref{OmegaAf} for transverse forces:
\begin{equation}\label{Omegav0f}
  \Omega_{v_0}f(\br^N,\buhat^N)\Biggr\rangle = \left(\Omega^\dagger_{-v_0}f(\br^N,\buhat^N)\right)\Biggr\rangle.
\end{equation}
We are interested in the evolution of the density mode $n(\bq)$:
\begin{equation}
  n(\bq,t) \equiv \sum_{i} \ee^{-i\bq\cdot\br_i(t)}
\end{equation}
and of its correlation function, the dynamical structure factor $S(\bq,t)$:
\begin{equation}
  S(\bq,t) \equiv \frac{1}{N}\left\langle n^*(\bq)n(\bq,t)\right\rangle.
\end{equation}
The following identity holds when inserted into an average:
\begin{equation}
  \partial_tn(\bq,t) = \Omega^\dagger_{v_0}n(\bq,t),
\end{equation}
so that $n(\bq,t) = \ee^{\Omega^\dagger_{v_0}}n(\bq,t)$. This, together with Eq.~\eqref{Omegav0f} allows us to write
\begin{equation}
  S(\bq,t) = \frac{1}{N}\left\langle n^*(\bq)\ee^{\Omega_{-v_0}t}n(\bq)\right\rangle.
\end{equation}
Another important quantity is the time derivative of the density mode at $t=0$:
\begin{equation}
  \partial_t n(\bq) = -iv_0q\hat u_\parallel(\bq),
\end{equation}
which is the projection of the velocity $\buhat_i$ along the wavevector $\bq$:
\begin{equation}
  \hat u_\parallel(\bq) \equiv \sum_i \left(\bqhat \cdot \buhat_i \right) \ee^{-i\bq\cdot\br_i}.
\end{equation}
We are also interested in the self-intermediate scattering function $F_s(\bq,t)$. Here
we re-write the definition \eqref{defFsq} by putting  the time dependence entirely into
$n_i(\bq,t)$,
\begin{equation}
  F_s(\bq,t) \equiv \frac{1}{N}\sum_{i=1}^N \left\langle n^*_i(\bq)n_i(\bq,t)\right\rangle.
\end{equation}
We have $\partial_t n_i(\bq) = -iv_0 qu_{\parallel,i}(\bq)$, with
\begin{equation}
  \hat u_{\parallel,i}(\bq) \equiv \bqhat \cdot \buhat_i \ee^{-i\bq\cdot\br_i}
\end{equation}
encoding the Fourier transform of the longitudinal self-propulsion velocity with respect to the wavevector $\bq$. The small-$q$ expansion of the self-intermediate scattering function yields the mean-squared displacement
\begin{equation}
  \Delta(t)\equiv \sum_{i=1}^N\left\langle \left[ \br_i(t) -\br_i(0)\right]^2\right\rangle.
\end{equation}
Using the isotropy of the dynamics given by Eq.~\eqref{eq:lABPs} we can also connect $F$ and $\Delta$:
\begin{equation}\label{F&DeltaABP}
  F_s(\bq,t) = 1 -\frac{q^2}{4}\Delta(t) + O(q^3).
\end{equation}
An equation for $\Delta$ is obtained at the end of the section.

\subsubsection{Short-time dynamics}

To gain insights on the dynamics we consider a short-time expansion of $S(\bq,t)$:
\begin{equation}\label{S(q,t)shortt}
  \begin{aligned}
    S(\bq,t) &\approx S(q) + \frac{t}{N}\left\langle n^*(\bq) \Omega_{-v_0} n(\bq)\right\rangle \\
    &+ \frac{t^2}{2N}\left\langle n^*(\bq) \Omega_{-v_0} \Omega_{-v_0} n(\bq)\right\rangle \\
    &= S(q) - iv_0q\left\langle n^*(\bq) \hat u_\parallel(\bq)\right\rangle \\
    &- \frac{t^2}{2}v_0^2q^2\left\langle \hat u^*_\parallel(\bq)\hat u_\parallel(\bq)\right\rangle .
  \end{aligned}
\end{equation}
The second term on the right hand side of Eq.~\eqref{S(q,t)shortt} vanishes due to rotational invariance. For the third term we have, using Einstein notation for the components of the vectors $\bqhat$ and $\buhat_i$,
\begin{equation}
  \begin{aligned}
    \left\langle \hat u^*_\parallel(\bq)\hat u_\parallel(\bq)\right\rangle &= \left\langle \sum_{i,j}\hat u_{i,\alpha}\hat u_{j,\beta}\hat q_\alpha \hat q_\beta \ee^{-i\bq\cdot\left(\br_i - \br_j\right)}\right\rangle \\
    &= \sum_{i,j}\left\langle \hat u_{\alpha,i}\hat u_{\beta,j}\right\rangle\left\langle \delta_{i,j}\delta_{\alpha\beta}\hat q_\alpha\hat q_\beta \ee^{-i\bq\cdot\left(\br_i-\br_j\right)}\right\rangle\\
    &= \frac{N}{2},
  \end{aligned}
\end{equation}
and therefore we obtain
\begin{equation}
  S(\bq,t) = S(q) - \frac{v_0^2q^2}{4}t^2 + O(t^3).
\end{equation}
This yields two results. First, we see that increasing $v_0$ produces a faster decay of the dynamic structure factor at short times. Moreover, in contrast with equilibrium overdamped Brownian dynamics, the decrease is quadratic in time (in line with the ballistic nature of the dynamics at short times), instead of being linear. This suggests that we should study the dynamics of $S(\bq,t)$ to second order in time, as done in the next section.

\subsubsection{Projection operator formalism}

We want to find an evolution equation for $S(\bq,t)$. The starting point is to write its second time derivative:
\begin{equation}\label{eq:partial2tS}
  \begin{aligned}
    \partial_t^2 S(\bq,t) &= \frac{1}{N}\left\langle n^*(\bq)\Omega_{-v_0}\ee^{\Omega_{-v_0}t}\Omega_{-v_0}n(\bq)\right\rangle \\
    &= -\frac{iv_0q}{N}\left\langle \hat u^*_\parallel(\bq)\Omega_{-v_0}\ee^{\Omega_{-v_0}t}n(\bq)\right\rangle .
  \end{aligned}
\end{equation}
By taking the Laplace Transform on both sides of Eq.~\eqref{eq:partial2tS} we get
\begin{equation}\label{eq:laplacepartial2tS}
  z^2S(\bq,z) - z = -\frac{iv_0q}{N}\left\langle \widehat u^*_\parallel(\bq)\Omega_{-v_0}\frac{1}{z-\Omega_{-v_0}}n(\bq) \right\rangle .
\end{equation}
We now have to choose a projection operator $\mP$ tailored to the relevant degrees of freedom. We assume these relevant modes to be the density field and its time derivative, which leads to the following expression for $\mathcal{P}$:
\begin{equation}
  \mathcal{P} \equiv \frac{1}{NS(q)}n(\bq)\rangle\langle n^*(\bq) + \frac{2}{N}\hat u_\parallel(\bq)\rangle\langle \hat u^*_\parallel(\bq).
\end{equation}
With this choice and a resolvent identity akin to Eq.~\eqref{eq:firstRid} previously used for transverse forces
\begin{equation}
    \begin{split}
        &\frac{1}{z-\Omega_{-v_0}} = \frac{1}{1-\mQ\Omega_{-v_0}\mQ} \\
        &+ \left(1+\frac{1}{z-\mQ\Omega_{-v_0}\mQ}\Omega_{-v_0}\right)\mP \frac{1}{z-\Omega_{-v_0}}\mP\\
        &\times\left(1+\Omega_{-v_0}\frac{1}{z-\mQ\Omega_{-v_0}\mQ}\right),
    \end{split}
\end{equation}
Eq.~\eqref{eq:laplacepartial2tS} becomes
\begin{equation}
  \begin{aligned}
    z^2S(\bq,z) &- zS(q) = -\frac{iv_0q}{N}\Biggl[\left\langle \hat u^*_\parallel(\bq) \Omega_{-v_0} n(\bq)\right\rangle \\
    &\times\frac{1}{NS(q)}\left\langle n^*(\bq)R(z)n(\bq)\right\rangle \\
    &+ \left\langle \hat u^*_\parallel(\bq) \Omega_{-v_0} \hat u_\parallel(\bq)\right\rangle \frac{2}{N}\left\langle \hat u^*_\parallel(\bq)R(z)n(\bq)\right\rangle \\
    &+ \left\langle \hat u^*_\parallel(\bq) \Omega_{-v_0} \mathcal{Q} R_{\mathcal{Q}}(z) \mathcal{Q} \Omega_{-v_0} \hat u_\parallel(\bq)\right\rangle \\
    &\times \frac{2}{N}\left\langle \hat u^*_\parallel(\bq)R(z)n(\bq)\right\rangle\Biggr].
  \end{aligned}
\end{equation}
An explicit computation gives:
\begin{equation}
  \begin{aligned}
    \left\langle \hat u^*_\parallel(\bq) \Omega_{-v_0} n(\bq)\right\rangle &= -i\frac{v_0qN}{2} , \\
    \left\langle \hat u^*_\parallel(\bq) \Omega_{-v_0} \hat u_\parallel(\bq)\right\rangle &= -\frac{N}{2}D_r , \\
    -\frac{iv_0q}{N}\left\langle \hat u^*_\parallel(\bq)\frac{1}{z-\Omega_{-v_0}}n(\bq)\right\rangle &= zS(\bq,z) - S(q).
  \end{aligned}
\end{equation}
Putting everything together and taking an inverse Laplace transform yields
\begin{equation}\label{eq:S_qt_lABP}
    \begin{split}
        \partial^2_tS(\bq,t) &= -\frac{v_0^2q^2}{2S(q)}S(\bq,t) - D_r\partial_tS(\bq,t) \\
        &- \int_0^{t}\dd \tau M(\bq,t-\tau)\partial_\tau S(\bq,\tau).
    \end{split}
\end{equation}
This is an evolution equation for the dynamical structure factor for the lifted-ABP dynamics. It allows for underdamped oscillations in the presence of a memory kernel given by
\begin{equation}
  M(\bq,t) \equiv -\frac{2}{N}\left\langle \hat u^*_\parallel(\bq) \Omega_{-v_0} \mathcal{Q} e^{\mathcal{Q}\Omega_{-v_0}\mathcal{Q}t} \mathcal{Q} \Omega_{-v_0} \hat u_\parallel(\bq)\right\rangle.
\end{equation}
With a similar approach, one can derive an equation for $F_s(\bq,t)$. In practice the step above must be repeated using the single particle projection operator 
\begin{equation}
    \mathcal{P}_i\equiv n_i(\bq)\rangle\langle n^*_i(\bq) + 2 u_{\parallel,i}(\bq)\rangle\langle \hat u^*_{\parallel,i}(\bq) ,
\end{equation}
instead of $\mathcal{P}$ and employ particle equivalence. The result is
\begin{equation}\label{eomFABP}\begin{split}
  \partial^2_t F_s(\bq,t) &= -\frac{v_0^2q^2}{2} F_s(\bq,t) - D_r\partial_t F_s(\bq,t) \\
  &- \int_0^{t}\dd \tau M_s(\bq,t-\tau)\partial_\tau F_s(\bq,\tau) ,
\end{split}\end{equation}
with the self-memory kernel defined as
\begin{equation}
  M_s(\bq,t) \equiv -\frac{2}{N}\sum_{i=1}^N \left\langle \hat u^*_{\parallel,i}(\bq) \Omega_{-v_0} \mathcal{Q} \ee^{\mathcal{Q}\Omega_{-v_0}\mathcal{Q}t} \mathcal{Q} \Omega_{-v_0} \hat u_{\parallel,i}(\bq)\right\rangle.
\end{equation}
All the steps performed so far are exact. To proceed, approximations are needed to compute $M$ and $M_s$. In the next section, we approximate the memory kernels using the mode-coupling scheme. 

\subsubsection{Mode-coupling expansion}

We expand the memory kernel as a product of density mode correlations. The physical justification for this choice is that the non-trivial part in the time evolution of $\hat u_\parallel(\bq)$ comes from the coupling with the fluid, which involves pairwise forces between particles. We therefore write
\begin{equation}\label{ABPmctexpM}\begin{split}
  \mathcal{Q}\Omega_{-v_0}\hat u_\parallel(\bq)\Biggr\rangle &= \frac{1}{2}\sum_\bk \frac{\left\langle n^*(\bk)n^*(\bq-\bk) \mathcal{Q}\Omega_{-v_0}\hat u_\parallel(\bq)\right\rangle}{N^2 S(k)S(\lvert\bq-\bk\rvert) } \\
  &n(\bk)n(\bq-\bk)\Biggr\rangle ,
\end{split}\end{equation}
with the factor $\frac{1}{2}$ inserted to avoid double counting of the density modes products. We now calculate the expectation value on the right hand side of Eq.~\eqref{ABPmctexpM} thanks to the convolution approximation $\left\langle n^*(\bk)n^*(\bq-\bk)n(\bq)\right\rangle \approx NS(q)S(k)S(\lvert\bq-\bk\rvert)$:
\begin{equation}
  \begin{aligned}
    & \left\langle n^*(\bk)n^*(\bq-\bk) \mathcal{Q}\Omega_{-v_0}\hat u_\parallel(\bq)\right\rangle \\
    &= \left\langle n^*(\bk)n^*(\bq-\bk)\Omega_{-v_0}\hat u_\parallel(\bq)\right\rangle \\
    &- \left\langle n^*(\bk)n^*(\bq-\bk) \mathcal{P}\Omega_{-v_0}\hat u_\parallel(\bq)\right\rangle \\
    &=  \left\langle \left[\Omega_{-v_0}^\dagger \left(n^*(\bk)n^*(\bq-\bk)\right)\right] \hat u_\parallel(\bq)\right\rangle \\
    &-  \left\langle n^*(\bk)n^*(\bq-\bk)n(\bq)\right\rangle\frac{1}{NS(q)}\left\langle n^*(\bq)\Omega_{-v_0}\hat u_\parallel(\bq)\right\rangle \\
    &= -iv_0 \bigg\langle \sum_{i,j,k}(\hat u_{i,\alpha}\hat q_\alpha)\left[\hat u_{j,\beta}k_\beta + \hat u_{k,\beta}(q_\beta - k_\beta) \right]\\
    &\times \ee^{i\bk\cdot\br_j}\ee^{i(\bq-\bk)\cdot\br_k}\ee^{-i\bq\cdot\br_i}\bigg\rangle + i\frac{Nv_0q}{2}S(k)S(\lvert\bq-\bk\rvert) \\
    &= i\frac{Nv_0}{2}\bigg[ q S(k)S(\lvert\bq-\bk\rvert) \\
    &- (\bqhat \cdot \bk)S(\lvert\bq-\bk\rvert) - \bqhat\cdot(\bq-\bk)S(k)\bigg] \\
    &= \
    i\frac{Nv_0\rho_0}{2}\left[ (\bqhat\cdot\bk)c(k) + \bqhat\cdot(\bq-\bk)c(\lvert\bq-\bk\rvert) \right] ,
  \end{aligned}
\end{equation}
where the direct correlation function  $\rho_0c(q)$ and the particle density of the system $\rho_0$ were defined below Eq.~\eqref{eq:tmctvertex} and above Eq.~\eqref{eq:phi_rpa}, respectively.

We can now give the expression for the mode-coupling memory kernel. To this end, we resort to the approximation
\begin{equation}
  \ee^{\mathcal{Q}\Omega_{-v_0}\mathcal{Q}t} \approx \ee^{\Omega_{-v_0}t},
\end{equation}
and use a Gaussian factorization for computing the two-body density correlation. Moreover, due to the irreversibility of the dynamics, we have
\begin{equation}
  \left\langle n^*(\bk)n^*(\bq-\bk) \mathcal{Q}\Omega_{-v_0}\hat u_\parallel(\bq)\right\rangle = \left\langle \hat u_\parallel(\bq) \Omega_{-v_0}\mathcal{Q}n(\bk)n(\bq-\bk)\right\rangle.
\end{equation}
The memory kernel reads, in the mode-coupling approximation
\begin{equation}\begin{split}
  M(\bq,t) &\approx \frac{v_0^2\rho_0}{4}\int\frac{d\bk}{(2\pi)^2} \\
  &\times\left[ (\bqhat\cdot\bk)c(k) + \bqhat\cdot(\bq-\bk)c(\lvert\bq-\bk\rvert)\right]^2 \\
  &\times S(\bk,t)S(\bq-\bk,t).
\end{split}\end{equation}
A similar expression can be obtained for $M_s$, using as a starting point the expansion
\begin{equation}\label{AOUPmctexpMs}\begin{split}
  \mathcal{Q}\Omega_{-v_0}\hat u_{\parallel,i}(\bq)\Biggr\rangle &= \sum_\bk \frac{\left\langle n^*(\bk)n_i^*(\bq-\bk) \mathcal{Q}\Omega_{-v_0}\hat u_{\parallel,i}(\bq)\right\rangle}{N S(k)} \\
  &\times n(\bk)n_i(\bq-\bk)\Biggr\rangle .
\end{split}\end{equation}  
The final result reads
\begin{equation}
  M_s(\bq,t) \approx \frac{v_0^2\rho_0}{2}\int\frac{d\bk}{(2\pi)^2}\left[\bqhat\cdot\bk c(k)\right]^2F(\lvert\bq-\bk\rvert,t)S(\bk,t).
\end{equation}
Note that the structure of the memory kernels $M$ and $M_s$ is the same as the one obtained for equilibrium dynamics, their magnitude being now determined by the magnitude of the self-propulsion speed $v_0$.

In the next sections we study the ergodicity breaking transition and the behavior of the diffusion constant stemming from these approximate expressions for the memory kernels. 

\subsubsection{Long-time dynamics: acceleration and ergodicity breaking}

To study the long-time dynamics we introduce the normalized dynamical structure function $\phi(\bq,t) \equiv \frac{S(\bq,t)}{S(q)}$. Its equation of motion reads
\begin{equation}\label{eq:phi_qt_lABP}
    \begin{split}
        \partial^2_t\phi(\bq,t) &= -\frac{v_0^2q^2}{2S(q)}\phi(\bq,t) - D_r\partial_t\phi(\bq,t) \\
        &- \int_0^{t}\dd \tau M(\bq,t-\tau)\partial_\tau \phi(\bq,\tau).
    \end{split}
\end{equation}
We first study the ergodicity breaking transition. We define the long-time limits of the normalized structure factor and the memory kernel
\begin{equation}
    \begin{split}
        \lim_{t\to+\infty} \phi(\bq,t) &= \phi_\infty(\bq) , \\
        \lim_{t\to+\infty} M(\bq,t) &= M_\infty(\bq).
    \end{split}
\end{equation}
In the long-time limit, Eq.~\eqref{eq:phi_qt_lABP} becomes a self-consistent equation for the plateau value of the correlation function:
\begin{equation}
    \begin{split}
        \frac{\phi_\infty(\bq)}{1-\phi_\infty(\bq)} &= \frac{2S(q)\rho_0}{q^2}\int\frac{d\bk}{(2\pi)^2} \\
        &\times\left[ (\bqhat\cdot\bk)c(k) + \bqhat\cdot(\bq-\bk)c(\lvert\bq-\bk\rvert)\right]^2 \\
        &\times \phi_\infty(\bk)\phi_\infty(\bq-\bk).
    \end{split}
\end{equation}
This equation is the same as the one obtained for overdamped equilibrium dynamics~\cite{szamel1991mode}. It follows that ergodicity breaking is expected to occur at the same critical density as in equilibrium.

To study the relaxation of the system in the ergodic regime, we introduce a relaxation time 
\begin{equation}
    \tau \equiv \int_0^{+\infty} \dd t \phi(\bq,t).
\end{equation}
Integration of Eq.~\eqref{eq:phi_qt_lABP} in $t$ from $0$ to $+\infty$ gives
\begin{widetext}
\begin{equation}\label{eq:ergodic_lABP}
    \tau = \frac{2S(q)D_r}{v_0^2q^2} + \frac{2S(q)\rho_0}{q^2}\int_0^{+\infty}\dd t\int\frac{d\bk}{(2\pi)^2}\left[ (\bqhat\cdot\bk)c(k) + \bqhat\cdot(\bq-\bk)c(\lvert\bq-\bk\rvert)\right]^2
    \phi(\bk,t)\phi(\bq-\bk,t). 
\end{equation}
\end{widetext}
This expression for $\tau$ is formally identical to the equilibrium case with the bare diffusion coefficient $D_0$ now being replaced with $\frac{v_0^2}{2D_r}=D_0$. While $D_0$ is slaved to the temperature in equilibrium, $\frac{v_0^2}{2D_r}$ can be independently varied in the lifted-ABP model, potentially leading to an accelerated dynamics. 

\subsubsection{Diffusion constant}

We obtain an equation of motion for the mean-squared displacement $\Delta(t)$ by substituting Eq.~\eqref{F&DeltaABP} in Eq.~\eqref{eomFABP} and retaining only the leading term in the limit $q\rightarrow 0$:
\begin{equation}\label{eomdeltaABP}
 \partial^2_t \Delta(t) = 2v_0^2 - D_r\partial_t\Delta(t) - \int_0^td\tau M_s(\bq\to 0, t-\tau)\partial_\tau\Delta(\tau).
\end{equation}
We now assume a diffusive behavior at large times: $\Delta(t)\sim 4Dt$. Substituting into Eq.~\eqref{eomdeltaABP} we obtain an expression for the diffusion constant $D$:
\begin{equation}
  D = \frac{v_0^2}{2D_r}\frac{1}{1 + \frac{v_0^2}{2D_r}m_{s}^\infty}, 
\end{equation}
where the memory kernel is given by
\begin{equation}
  m_{s}^{\infty} = \rho_0\int_0^{+\infty}\dd t\int\frac{d\bk}{(2\pi)^2}\left[\bqhat\cdot\bk c(k)\right]^2 F_s(k,t)S(k,t).
\end{equation}
The formal expression of $m_{s}^\infty$ is the same as in equilibrium, while the one for $D$ is also formally similar, with the replacement $D_0$ by $\frac{v^2_0}{2D_r}$. In the ergodic phase when $m_s^\infty$ is finite, the diffusion constant $D$ can be enhanced with respect to equilibrium by increasing $\frac{v^2_0}{2D_r}>D_0$, which plays the role of the driving force. Differently from the transverse force case, the acceleration survives the high-temperature limit, since  $D=\frac{2v_0^2}{D_r}$ in this limit. In the opposite limit of temperature approaching the dynamic transition, $m_{s,\infty}$ diverges and thus $D$ exactly recovers its equilibrium expression and the lifted-ABP dynamics is no longer felt. This suggests that lifted dynamics acceleration plummets as the kinetic glass transition is approached. This is in line with recent numerical results~\cite{ghimenti2024irreversible}. 

\section{Conclusion}

\label{sec:outlook}

The irreversible sampling of the Boltzmann distribution can generically be faster than a conventional equilibrium dynamics. Our goal in this work was to study theoretically the acceleration and the microscopic relaxation dynamics obtained in the case of dense liquids approaching the glass transition when irreversible dynamics are used~\cite{ghimenti2022accelerating,ghimenti2023sampling,ghimenti2024irreversible}. Unlike our precedent effort~\cite{ghimenti2024transverse} where the problem was analysed exactly in the mean-field limit of liquids in large dimensions, we developed here an approximate mode-coupling approach to study the dynamics directly in finite dimension. These two approaches are thus complementary~\cite{parisi2020theory}. 

To this end, we developed a novel mode-coupling approach specifically tailored to address nonreciprocal transverse forces, which can be seen as one of the simplest nonequilibrium drive achieving acceleration. Technically, this has required a careful treatment of transverse and longitudinal modes. Physically, we concluded that the location of the ergodicity breaking transition predicted to occur in equilibrium without transverse forces is never affected by transverse forces.  However, the dynamics in the ergodic phase is systematically accelerated by transverse forces, with a non-trivial temperature dependence. These results are fully consistent with both the large dimensional results~\cite{ghimenti2024transverse} and computer simulations~\cite{ghimenti2023sampling}. 

The emerging physical picture is that longitudinal diffusion gets arrested for the same value of the control parameters as in equilibrium dynamics, but the energy injected by the transverse forces is dissipated in an ever-going odd motion, that we picture as a swirling motion within an environment that becomes increasingly confining as the temperature is lowered. Approaching the kinetic glass transition, the odd diffusion saturates to a constant value, the odd mobility vanishes, and the odd viscosity diverges.

In a recent work~\cite{ghimenti2023sampling} we further argued that transverse forces share the same accelerating physical ingredient as their lifted counterparts in which the nonequilibrium drive results from coupling the dynamics to extra degrees of freedom. Our explicit mode-coupling approach clearly supports this intuition. For the lifted processes explicitly considered in our final section, the physical conclusions echo those drawn from considering transverse forces. We believe that this analogy is a credible explanation of the observations of \cite{ghimenti2024irreversible} showing that event-chain Monte Carlo methods lose their edge over conventional Metropolis Monte Carlo as the fluid enters more deeply into its glassy regime. 

Our work opens at least two immediate research directions. The first one has to do with the collective behavior of dense assemblies of chiral active particles~\cite{liebchen2022chiral,debets2023glassy}, for which a mode-coupling analysis will surely closely follow the footsteps developed in our work. The second one is more subtle: it is concerned with the numerical integration of the derived mode-coupling equations. This will surely be a highly nontrivial task due to the loss of isotropy. The absence of a fluctuation-dissipation theorem is additionally responsible for increasing the number of coupled self-consistent integro-differential equations to solve, which will add to the complexity of the numerical solution. 
 
\begin{acknowledgments}

LB, FG and FvW acknowledge the financial support of the ANR THEMA AAPG2020 grant. GS acknowledges the support of NSF Grant No.~CHE 2154241.

\end{acknowledgments}

\appendix

\section{Irreducible memory kernel}

In this appendix we justify the choice for the irreducible memory kernel made in Eq.~\eqref{eq:Omegairr}. We first review how the irreducible memory kernel is introduced in the equilibrium case, when $\gamma=0$. Following Kawasaki~\cite{kawasaki1995irreducible}, the starting point is the observation that the evolution operator $\Omega \equiv D_0 \sum_i \bnabla_i \left[ \beta \bF_i + \bnabla_i \right]$ can be mapped to a Hermitian operator $H \equiv \ee^{\frac{\beta \mathcal{H}}{2}}\Omega \ee^{\frac{-\beta \mathcal{H}}{2}}$. It can be written in a manifestly Hermitian form as
\begin{equation}
    H = -\sum_i \bU_i^\dagger \cdot \bU_i ,
\end{equation}
with $\bU_i \equiv \sqrt{D_0}\left[ -\bnabla_i + \frac{\beta}{2}\bF_i\right]$. Going back to the non-Hermitian representation we see that
\begin{equation}
    \Omega = -\sum_i \ee^{\frac{-\beta \mathcal{H}}{2}}\bU_i^\dagger \ee^{\frac{\beta \mathcal{H}}{2}}\cdot \ee^{\frac{-\beta \mathcal{H}}{2}}\bU_i\ee^{\frac{\beta \mathcal{H}}{2}} \equiv -\sum_i \bO^{\times}_i\cdot \bO_i ,
\end{equation}
with 
\begin{equation}
    \begin{split}
        \bO_i^\times &= \sqrt{D_0}\bnabla_i , \\
        \bO_i &= \sqrt{D_0}\left[ -\bnabla_i +  \beta \bF_i\right].
    \end{split}
\end{equation}
At this point, one can insert any generic projection operator $\mathcal{P}_i$ and its orthogonal counterpart $\mathcal{Q}_i\equiv \mathcal{I} - \mathcal{P}_i$. For consistency with the notation of the main text, we also enclose $\Omega$ between the orthogonal projector $\mathcal{Q}=\mathcal{I} - \mathcal{P}$, defined  from Eq.~\eqref{eq:P}:
\begin{equation}
    \begin{split}
        \mathcal{Q}\Omega\mathcal{Q} &= -\mathcal{Q} \left(\sum_i \bO^\times_i \mathcal{P}_i \cdot \bO_i -\sum_i \bO^\times_i \mathcal{Q}_i \cdot \bO_i\right)\mathcal{Q} \\
        &\equiv -\delta\Omega + \Omega^{\text{irr}}  .
    \end{split}
\end{equation}
What remains to do is to properly choose the projection operator $\mathcal{P}$, in such a way that a renormalization of the memory kernel occurs: $\widetilde M(q,z) = \frac{M(q,z)}{1 + c M(q,z)}$, with $c$ some $q$-dependent constant. Following Cichocki and Hess~\cite{cichocki1987memory} we take
\begin{equation}\label{eq:Pi}
    \mathcal{P}_i \equiv \ldots \ee^{-i\bq\cdot\br_i}\rangle \langle \ee^{i\bq\cdot\br_i}\ldots ,
\end{equation}
leading to the following expression for $\delta\Omega$
\begin{equation}
    \delta\Omega = \frac{D_0 \beta^2 }{N} \ldots\mathcal{Q}\bff(\bq)\rangle\cdot\langle\bff^*(\bq)\mathcal{Q}\ldots
\end{equation}  
Note that this is different from the choice of Kawasaki, which is instead
\begin{equation}
    \mathcal{P}_j^{\text{Kawasaki}} = \ldots \bO_j \ee^{-i\bq\cdot\br_j}\rangle \frac{1}{D_0 q^2}\langle \ee^{i\bq\cdot\br_j}\bO^\times_j\ldots
\end{equation}
which leads to, using the fact that particles are equivalent,
\begin{equation}
    \begin{split}
        \delta\Omega^{\text{Kawasaki}} &= \ldots \Omega n(\bq)\rangle\frac{1}{\langle n^*(\bq) \Omega n(\bq)\rangle} \langle n^*(\bq)\Omega\ldots \\
        &= \frac{D_0\beta^2}{N} \ldots \mathcal{Q} f_\parallel(\bq) \rangle \langle f_\parallel^*(\bq)\mathcal{Q}\ldots
    \end{split}
\end{equation}
We see that in equilibrium, the operator $\delta\Omega^{\text{Kawasaki}}$  contains only contributions from the longitudinal currents.

We now turn to the general situation with $\gamma\neq 0$. In this case one can see that an analogous decomposition for $\Omega_\gamma$ can be made, with an extra term to take into account the presence of transverse forces:
\begin{equation}
    \Omega_{-\gamma} = -\bO^\times \cdot \left(\mathbf{1} - \gamma \bA\right) \cdot \bO .
\end{equation}
Using the projection operator $P_i$ defined in Eq. \eqref{eq:Pi} we obtain
\begin{equation}
    \begin{split}
        \Omega^{\text{irr}}_{-\gamma} &= \mathcal{Q} \bO^\times \mathcal{Q}_i \cdot \left(\mathbf{1} - \gamma \bA\right) \cdot \bO \\
        &= D_0 \mathcal{Q}\sum_j \bnabla_j \mathcal{Q}_j \cdot \left(\mathbf{1} - \gamma\bA \right) \left[ -\beta \bF_j + \bnabla_j\right] \mathcal{Q} ,
    \end{split}
\end{equation}
which is Eq. \eqref{eq:Omegairr} shown in the main text.

\bibliography{final_biblio}

\end{document}